\documentclass[aps,prd, twocolumn,superscriptaddress,floatfix,preprintnumbers,nofootinbib]{revtex4-1}
\usepackage[utf8]{inputenc}
\usepackage{xspace}
\usepackage{amsmath}
\usepackage{amssymb}
\usepackage{bm}
\usepackage{xcolor}
\usepackage{feynmf}
\usepackage{slashed}
\usepackage{braket}
\usepackage{url}
\usepackage{natbib}
\usepackage{graphicx}
\usepackage{subfigure}
\usepackage{mathrsfs}  
\usepackage{cancel}
\usepackage[normalem]{ulem}
\usepackage[pdftex,bookmarks,linktocpage,pdfpagelabels,plainpages=false,hyperfigures,linkcolor=blue,citecolor=blue]{hyperref}
\hypersetup{colorlinks=true}
\newcommand{\red}[1]{\textcolor{red}{#1}}

\begin{document}

\title{Neutrino Nonstandard Interactions from $LLE$-type $R$-parity Violation}

\author{Fang~Xu}
\email{xufang@wustl.edu}
\affiliation{Department of Physics, Center for Field Theory and Particle Physics, Fudan University, Shanghai, 200433, China}
\affiliation{Department of Physics and McDonnell Center for the Space Sciences,
Washington University, St.~Louis, MO 63130, USA}

\begin{abstract}
We study neutrino nonstandard interactions (NSI) induced by the trilinear $LLE$ couplings in $R$-parity-violating supersymmetry in a split-family spectrum where third-generation sleptons can be comparatively light. We systematically classify the relevant coupling patterns, derive the corresponding effective interactions and NSI parameters, and confront the resulting parameter space with a broad set of experimental constraints. We find that percent-level NSI can be realized in selected channels, with the largest magnitudes reaching $\varepsilon_{ee}\simeq 8\%$ and $\varepsilon_{\mu\mu}\simeq 3.8\%$, while other entries are typically restricted to the subpercent level (including $\varepsilon_{\mu\tau}\simeq -0.3\%$ in the viable off-diagonal case). We also highlight the structural origin of the different constraint patterns between $\widetilde{\tau}_R$- and $\widetilde{\tau}_L$-mediated scenarios and comment on the relation to the Zee model.
\end{abstract}

\maketitle

\section{Introduction} \label{sec:intro}

Neutrino oscillation experiments have established that neutrinos are massive and undergo flavor mixing~\cite{Super-Kamiokande:1998kpq,SNO:2002tuh,KamLAND:2002uet}, thereby providing a clean window into physics beyond the Standard Model (SM)~\cite{Davidson:2003ha,Hernandez:2009fr,Ohlsson:2012kf}. In terrestrial and astrophysical environments, neutrino propagation is additionally shaped by coherent forward scattering in matter, which in the SM is encoded in the Mikheyev–Smirnov–Wolfenstein (MSW) potential~\cite{Wolfenstein:1977ue,Mikheyev:1985zog}. In the original formulation of the matter effect, it was emphasized that new interactions involving neutrinos could modify the effective potential and consequently distort oscillation probabilities in a characteristic way~\cite{Wolfenstein:1977ue,Mikheyev:1985zog,Valle:1987gv,Roulet:1991sm,Guzzo:1991hi}. Motivated by this observation, the concept of neutrino nonstandard interactions (NSI) encapsulates the possibility that neutrinos may interact with matter in ways that go beyond the standard weak interactions predicted by the SM. Such effects are conveniently parametrized by NSI parameters~\cite{Grossman:1995wx,Antusch:2008tz,Miranda:2015dra,Farzan:2017xzy,Proceedings:2019qno}, which capture a broad class of new physics scenarios in a compact and largely model-independent way through their deviations from the standard SM matter potential.

At energies well below the mediator scale, NSI can be parametrized by effective four-fermion contact interactions between neutrinos and charged fermions. In this low-energy description, one supplements the SM Lagrangian with additional terms encoding possible nonstandard neutrino–matter interactions. For neutral-current (NC) and charged-current (CC) NSI relevant to neutrino propagation in matter, we use the following effective parametrization~\cite{Wolfenstein:1977ue,Berezhiani:2001rs,Davidson:2003ha,Ohlsson:2012kf,Farzan:2017xzy,Babu:2019mfe}
{\fontsize{9.5}{12}\selectfont
\begin{align}
&\mathcal{L}_{\mathrm{NSI}}^{\mathrm{NC}} = -2 \sqrt{2} G_{F} \sum_{f, X, \alpha, \beta} \varepsilon_{\alpha \beta}^{f X}\left(\bar{\nu}_{\alpha} \gamma^{\mu} P_{L} \nu_{\beta}\right)\left(\bar{f} \gamma_{\mu} P_{X} f\right)~,\label{eq:NC NSI}\\
&\mathcal{L}_{\mathrm{NSI}}^{\mathrm{CC}} = -2 \sqrt{2} G_{F} \sum_{f, f^{\prime}, X, \alpha, \beta} \varepsilon_{\alpha \beta}^{f f^{\prime} X}\left(\bar{\nu}_{\alpha} \gamma^{\mu} P_{L} \ell_{\beta}\right)\left(\bar{f}^{\prime} \gamma_{\mu} P_{X} f\right)~,\label{eq:CC NSI}
\end{align}
}
where $X=L,R$ and $f,f^\prime \in \{ e,u,d \}$. The coefficients $\varepsilon_{\alpha \beta}$ in Eqs.~\eqref{eq:NC NSI} and~\eqref{eq:CC NSI} are dimensionless parameters that quantify the size of NSI relative to the SM weak interaction strength, normalized to the Fermi constant $G_F$. They encode the flavor structure of the interaction through diagonal versus flavor-changing entries, and off-diagonal phases may introduce additional $CP$ violation. In this way, the set of NSI parameters provides a compact characterization of how new physics could modify neutrino interactions with matter at low energies.

For neutrino propagation in ordinary matter, the relevant matter potential arises from coherent forward scattering of neutrinos on the background medium and is therefore governed by NC interactions of the form $(\bar{\nu}_{\alpha} \gamma^{\mu} P_{L} \nu_{\beta})(\bar{f} \gamma_{\mu} P_{X} f)$. CC operators instead involve $\nu \to \ell$ transitions and typically affect neutrino production and detection processes rather than the propagation potential. Consequently, when discussing matter effects we focus on the NC NSI in Eq.~\eqref{eq:NC NSI}. In this effective description, one can further note that matter effects are controlled by the vector part of the background fermion current. This is because the medium is, to an excellent approximation, electrically neutral, nonrelativistic, and unpolarized, so that the expectation value of the axial current vanishes, $\langle \bar{f} \gamma^\mu \gamma^5 f \rangle \simeq 0$, while the time component of the vector current simply counts number density, $\langle \bar{f} \gamma^0 f \rangle = N_f$. As a consequence, the effective matter potential induced by NSI depends on the vector combination of left- and right-handed couplings,
\begin{equation}
    \varepsilon_{\alpha\beta}^{fV} = \varepsilon_{\alpha\beta}^{fL}+\varepsilon_{\alpha\beta}^{fR}~.
\end{equation}
In the flavor basis, this potential enters the propagation Hamiltonian as an additive term proportional to the electron density, with the relative contributions of other fermions determined by their abundances in the medium. It is therefore convenient to define the effective NSI parameter relevant for oscillations as the density-weighted superposition
\begin{equation}
\begin{aligned}
\varepsilon_{\alpha \beta} &=\sum_{f \in\{e, u, d\}}\frac{N_{f}(x)}{N_{e}(x)} \varepsilon_{\alpha \beta}^{f V} \\
&=\varepsilon_{\alpha \beta}^{e V}+\frac{N_{p}(x)}{N_{e}(x)}\left(2 \varepsilon_{\alpha \beta}^{u V}+\varepsilon_{\alpha \beta}^{d V}\right)+\frac{N_{n}(x)}{N_{e}(x)}\left(\varepsilon_{\alpha \beta}^{u V}+2 \varepsilon_{\alpha \beta}^{d V}\right)
\\
&=\varepsilon_{\alpha \beta}^{e V}+\left[2+Y_{n}(x)\right] \varepsilon_{\alpha \beta}^{u V}+\left[1+2 Y_{n}(x)\right] \varepsilon_{\alpha \beta}^{d V}~,
\end{aligned}
\end{equation}
where $N_f(x)$ denotes the number density of fermion $f$ along the neutrino trajectory and $Y_{n}(x) \equiv N_{n}(x) / N_{e}(x)$. In the second line, we have used the quark content of nucleons, $p=(uud)$ and $n=(udd)$. In the last line, we imposed electric charge neutrality, $N_p(x) = N_e(x)$, of the medium. When the medium composition is approximately constant, the explicit $x$-dependence can be dropped; otherwise one may interpret the above expression locally along the trajectory.

NSI parameters provide a compact phenomenological language to encode possible new physics effects in neutrino interactions with matter~\cite{Davidson:2003ha,Hernandez:2009fr,Ohlsson:2012kf,Proceedings:2019qno,Farzan:2017xzy,Miranda:2015dra,Liao:2016hsa,Bischer:2018zbd,Babu:2019mfe,Babu:2020nna,Dev:2023znd}. They may originate from heavy mediators integrated out at low energies, or from scenarios that violate lepton flavor and induce flavor off-diagonal interactions in the neutrino sector. In this way, the NSI parameters offer a practical bridge between oscillation observables and microscopic model building, and allow one to assess how NSI could modify propagation in matter and the interpretation of precision oscillation data~\cite{NOvA:2021nfi,T2K:2023smv,2852068,Esteban:2024eli}. Experimentally, current and upcoming neutrino programs~\cite{DUNE:2015lol,DUNE:2016hlj,DUNE:2016rla,Hyper-Kamiokande:2018ofw,JUNO:2015zny} aim to test such effects through increasingly precise measurements of oscillation patterns across different baselines and matter environments, complemented by probes of neutrino scattering and other signatures of nonstandard dynamics.

The purpose of this paper is to study NSI induced by $R$-parity violating (RPV) supersymmetry, specializing to the $LLE$ sector. Concretely, we examine which RPV couplings $\lambda_{ijk}$ can generate sizable NSI, how they map onto the effective NSI parameters, and what ranges remain compatible with existing constraints. Our analysis is carried out in the “RPV3” framework, i.e. a minimal third-generation-dominated RPV setup motivated by naturalness~\cite{Papucci:2011wy,Feng:2013pwa,Altmannshofer:2017poe} and by the stringent constraints from flavor- and $CP$-violating observables, which in generic supersymmetric (SUSY) spectra tend to favor either near-degeneracy/alignment or a hierarchical (split-family) setup with heavier first- and second-generation sfermions~\cite{Dimopoulos:1995ju,Gabbiani:1996hi,Cohen:1996vb,Agashe:1998zz,Hisano:2000wy,Giudice:2008uk}. In this framework, the first two generations of sfermions are decoupled, while the third-generation sleptons/sneutrinos can remain near the TeV scale, allowing potentially visible low-energy imprints without conflicting with stringent flavor data~\cite{Altmannshofer:2020axr,BhupalDev:2021ipu,Afik:2022vpm}. Finally, we stress the scope of the present study. The assumption that only the $LLE$ interactions are nonzero is adopted here mainly as a phenomenological benchmark. From the theoretical side, this setup avoids the baryon-number-violating $UDD$ operators.\footnote{Proton decay is only induced at an unacceptable level when both baryon- and lepton-number violating RPV operators are present simultaneously. Therefore, allowing $\lambda$ and $\lambda^\prime$ is compatible with proton-decay constraints as long as the baryon-number violating $UDD$ $\lambda^{\prime\prime}$ couplings are sufficiently small~\cite{Martin:1997ns,Barbier:2004ez}, which can be ensured by the baryon triality~\cite{Ibanez:1991hv,Ibanez:1991pr}.} As for the further suppression of the lepton-number-violating $LQD$ interactions, hierarchical patterns among the RPV couplings can in principle arise from suitable flavor-dependent symmetries or UV textures~\cite{Carone:1996nd,Monteux:2013mna}, although we do not attempt to construct an explicit underlying model here. Phenomenologically, if additional RPV operators such as $LQD$ are switched on simultaneously, they typically introduce further, often unavoidable, constraints from flavor, charged-lepton, and collider processes, which tend to reduce the allowed NSI size. The present $LLE$-only setup therefore corresponds to a best-case scenario, and our results should be read as robust upper envelopes for NSI that can be attributed to $LLE$ interactions in an RPV3-like spectrum.

The rest of the paper is organized as follows. In Sec.~\ref{sec:frame}, we briefly introduce the RPV3 framework and classify the relevant $LLE$ interactions into several representative cases for inducing sizable NSI. In Secs.~\ref{sec:parameters} and~\ref{sec:constraints}, we derive the corresponding relations to the effective NSI parameters and discuss the relevant theoretical and experimental constraints. In Sec.~\ref{sec:numerical}, we present a numerical analysis and summarize the resulting parameter ranges for NSI. The conclusions are given in Sec.~\ref{sec:conclusions}.

\section{RPV3 Setup} \label{sec:frame}

As discussed in Sec.~\ref{sec:intro}, the RPV3 setup provides a simple way to accommodate stringent flavor and $CP$ constraints without abandoning the naturalness-motivated preference for comparatively light third-generation superpartners. Accordingly, the first two generations of sfermions are taken to be heavy and effectively decoupled, while the third-generation sfermions can remain around the TeV scale~\cite{Altmannshofer:2020axr,BhupalDev:2021ipu,Afik:2022vpm}. In the following, we focus on the $LLE$ sector, in which the NSI effects considered in this work are mediated by third-generation leptonic sfermions. The RPV interactions considered in this work are described by the trilinear $LLE$ term in the superpotential,
\begin{equation} \label{eq:W_RPV}
    W_\text{RPV} = \frac{1}{2} \lambda_{ijk} \hat L_i \hat L_j \hat E^c_k ~,
\end{equation}
where $\hat L$ and $\hat E^c$ are the left-handed lepton doublet and right-handed charged-lepton singlet superfields, respectively. The $LLE$ couplings are antisymmetric in the first two indices, $\lambda_{ijk} = - \lambda_{jik}$, and therefore $\lambda_{iik} = 0$, with $i,j,k$ labeling flavors.

Expanding Eq.~\eqref{eq:W_RPV} in component fields, one obtains the lepton-number-violating interactions between leptons and sleptons,
\begin{align} \label{eq:RPVLLE}
    \mathcal L_{LLE} \supset - \frac{1}{2}\lambda_{ijk}\big[ &\widetilde{\nu}_{iL} \bar{e}_{kR} e_{jL} +\widetilde{e}_{jL} \bar{e}_{kR}\nu_{iL} +\widetilde{e}_{kR}^{*} \bar{\nu}_{iL}^c e_{jL}\nonumber\\
    &- (i\leftrightarrow j) \big]+{\rm H.c.} ~,
\end{align}
where $e_{L,R}$ and $\nu_L$ are the charged-lepton and neutrino fields, and $\tilde e_{L,R}$, $\tilde\nu_L$ denote the corresponding scalar superpartners. The above Lagrangian contains $3^2=9$ independent $\lambda_{ijk}$ couplings. In the RPV3 framework, where the first two generations of sfermions are decoupled, the number of $LLE$ couplings relevant for our analysis is reduced to seven. Among these couplings, we summarize in Table~\ref{tab:LLE} the $\lambda$ interactions in the RPV3 setup that can induce neutrino NSI. For NSI in ordinary matter, the effective interaction must involve electrons, which requires at least one index to be 1; meanwhile, the RPV3 assumption implies that another index should be 3 so that the exchange of a third-generation slepton is involved. The red entries in Table~\ref{tab:LLE} indicate the interactions that can induce NSI, whereas the remaining entries typically lead instead to additional phenomenological constraints that are unavoidable in each case.

In this context, it is useful to comment on the relation to the Zee model~\cite{Zee:1980ai}, which has been discussed as a representative setting for NSI-related phenomenology~\cite{Babu:2019mfe}. At the level of flavor structure, the $LLE$ interaction shares the same antisymmetric $L_iL_j$ contraction as the Zee-type Yukawa coupling $f_{ij}\,L_i^T C i\sigma_2 L_j\,h^+$, which may suggest a close correspondence. However, the two frameworks differ in an important and model-specific way. In the supersymmetric case, the single superpotential term $\frac{1}{2}\,\lambda_{ijk}\,\hat L_i\hat L_j\hat E^c_k$ necessarily generates, upon component expansion, several distinct fermion--sfermion interactions involving different scalar superpartners [schematically $\tilde\nu_{iL}\,\bar e_{kR}e_{jL}$, $\tilde e_{jL}\,\bar e_{kR}\nu_{iL}$, and $\tilde e_{kR}^*\,\bar{\nu}_{iL}^ce_{jL}$, together with $(i\leftrightarrow j)$ terms], all controlled by the same $\lambda_{ijk}$ parameter. Consequently, once a given $\lambda_{ijk}$ is switched on, the accompanying interaction terms cannot be removed independently: even if only a subset of them contributes to NSI, the remaining terms typically induce additional phenomenological constraints, whose impact is spectrum-dependent.

In particular, the $LLE$ interaction involves several scalars as mediators, so the relative strength of the correlated constraints depends on the corresponding mass parameters and may be reduced if some of these states are sufficiently heavy. By contrast, the Zee model introduces a single charged scalar $h^+$ with the Yukawa structure $f_{ij}L_iL_j h^+$, and does not entail the supersymmetric correlation among multiple sfermion-mediated vertices tied to one coupling. As a result, despite superficial similarities at the level of the interaction structure, the mapping to the NSI parameters $\varepsilon_{\alpha\beta}$ and the correlated constraint pattern in the $LLE$ case can differ significantly from those in the Zee scenario, which motivates the analysis of $LLE$-induced NSI in the RPV3 framework presented below.

{\renewcommand{\arraystretch}{1.8}
\begin{table}[htbp]
\caption{Possible $LLE$ couplings relevant for neutrino NSI in the RPV3 framework. The red-highlighted interaction terms could induce NSI; the other terms tied to the same $\lambda_{ijk}$ couplings imply additional constraints. The couplings shown in the table are selected based on the requirement of an electron leg in matter and third-generation slepton exchange in RPV3, leading to one first- and one third-generation index.}
\begin{tabular}{c|c|c|c|c}
\hline
$\lambda_{123}$                                                                & $\lambda_{131}$                                                 & $\lambda_{132}$                                                    & $\lambda_{133}$                                                                                      & $\lambda_{231}$                                                  \\ \hline \hline
$\widetilde{\tau } _{R}^{\ast } \overline{\nu } _{eL}^{c} \mu _{L}$   & \red{$\widetilde{\tau } _{L} \overline{e} _{R} \nu _{eL}$}    & $\widetilde{\tau } _{L} \overline{\mu } _{R} \nu _{eL}$   & $\widetilde{\tau } _{L} \overline{\tau } _{R} \nu _{eL}$                                    & \red{$\widetilde{\tau } _{L} \overline{e} _{R} \nu _{\mu L}$}  \\ \hline
\red{$- \widetilde{\tau } _{R}^{\ast } \overline{\nu } _{\mu L}^{c} e _{L}$} & $- \widetilde{\nu } _{\tau L} \overline{e} _{R} e_{L}$ & $- \widetilde{\nu } _{\tau L} \overline{\mu } _{R} e_{L}$ & $\widetilde{\tau } _{R}^{\ast } \overline{\nu } _{eL}^{c} \tau _{L}$                        & $- \widetilde{\nu } _{\tau L} \overline{e} _{R} \mu _{L}$ \\ \hline
                                                                               &                                                                 &                                                                    & $- \widetilde{\nu } _{\tau L} \overline{\tau } _{R} e_{L}$                                  &                                                                  \\ \hline
                                                                               &                                                                 &                                                                    & \red{$- \widetilde{\tau } _{R}^{\ast } \overline{\nu } _{\tau L}^{c} e _{L}$} &                                                                  \\ \hline
\end{tabular}
\label{tab:LLE}
\end{table}
}

From Table~\ref{tab:LLE}, one can see that the interactions associated with $\lambda_{132}$ do not contribute to neutrino NSI. Within the scope of this work, we therefore identify six representative cases that can induce neutrino NSI: (a) $\lambda_{123}\neq 0$, (b) $\lambda_{131}\neq 0$, (c) $\lambda_{133}\neq 0$, (d) $\lambda_{231}\neq 0$, (e) $\lambda_{123},\lambda_{133}\neq 0$, and (f) $\lambda_{131},\lambda_{231}\neq 0$, where cases (a)--(d) are treated under the single-coupling dominance hypothesis, cases (e) and (f) are treated under the double-coupling dominance hypothesis, and all other $LLE$ couplings are taken to be zero in each case. In each case, only the interaction terms highlighted in red contribute to the NSI parameters.

A comment is in order for case (f). Although the combination $\lambda_{131}$ and $\lambda_{231}$ can, in principle, generate an off-diagonal NSI parameter $\varepsilon_{e\mu}$, the allowed effect is rendered negligible by the very stringent charged-lepton flavor-violating (cLFV) bound from $\mu\to ee\bar e$. In our setup, this decay receives a tree-level contribution from the $LLE$ interactions in Table~\ref{tab:LLE} via $\widetilde{\nu}_\tau$ exchange, which strongly suppresses the viable parameter space for $\varepsilon_{e\mu}$ in case (f). The model-independent lower bound on $m_{\widetilde{\nu}_\tau}$ is $41~\mathrm{GeV}$, obtained from the LEP measurement of the invisible $Z$-boson decay width~\cite{ALEPH:1991qhf}. In contrast, typical limits from multilepton searches for $\widetilde{\nu}_\tau$ pair production sets $m_{\widetilde{\nu}_\tau}$ values of $\mathcal{O}(\mathrm{TeV})$~\cite{ATLAS:2018rns,BhupalDev:2021ipu,Afik:2022vpm}. For a $\widetilde{\nu}_\tau$ with mass $m_{\widetilde{\nu}_\tau} = 1$~TeV, the bound on the product of couplings in case (f) is $\lambda_{131}\lambda_{231}<6.6\times 10^{-5}$~\cite{Dreiner:2006gu}, which implies an off-diagonal NSI parameter $\varepsilon_{e\mu}\lesssim \mathcal{O}(10^{-6})$ for $m_{\widetilde{\tau}_L}\sim \mathcal{O}(\mathrm{TeV})$. We therefore do not delve further into the analysis of case (f) in this work.

The cases (a)--(e) are not arbitrary single- or double-coupling examples, but representative configurations that capture the phenomenologically distinct outcomes relevant to the present analysis. More generally, for $LLE$ coupling configurations within the RPV3 framework involving one, two, or more dominant couplings, the resulting NSI pattern either reduces to the representative cases (a)--(e), or else becomes negligibly small because the additional interactions are subject to strong constraints, similar to case (f).
Finally, throughout the remainder of this paper, we adopt the following convention for simplicity that the labels (a), (b), (c), (d), and (e) used in equations, figures, and tables refer to the corresponding cases (a), (b), (c), (d), and (e) defined above.

\begin{figure*}[htbp]
    \centering
    \subfigure[$\ \lambda_{123}\neq 0$]{\includegraphics[width=3.5cm]{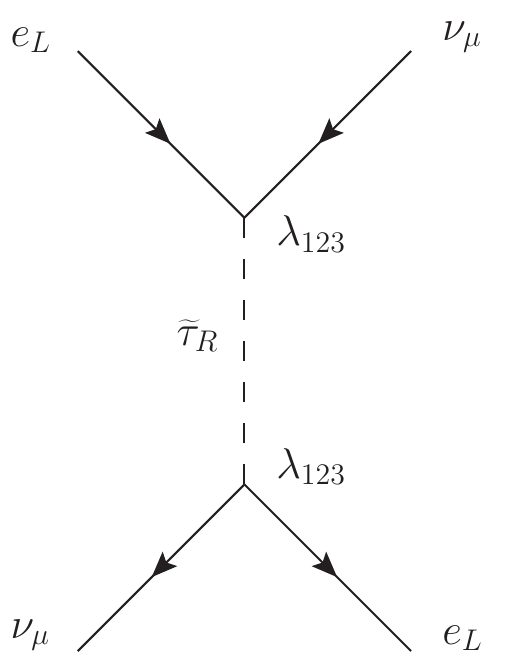}}
    \subfigure[$\ \lambda_{131}\neq 0$]{\includegraphics[width=3.5cm]{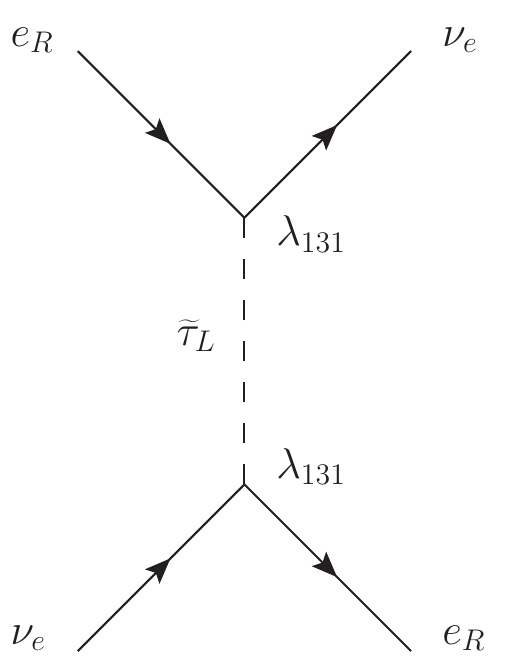}}
    \subfigure[$\ \lambda_{133} \neq 0$]{\includegraphics[width=3.5cm]{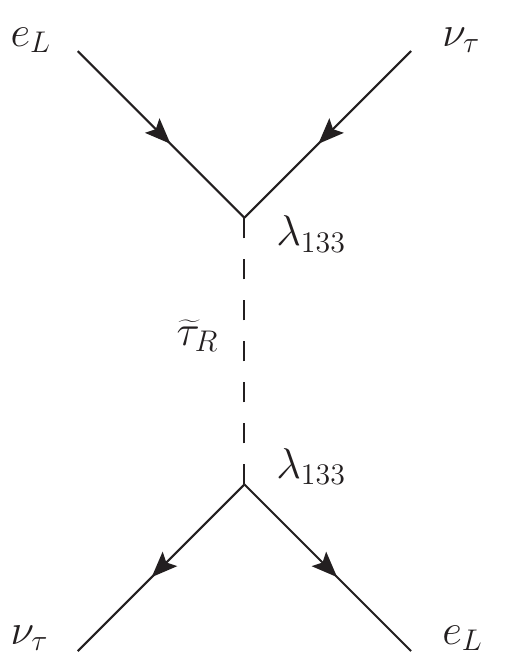}}
    \subfigure[$\ \lambda_{231}\neq 0$]{\includegraphics[width=3.5cm]{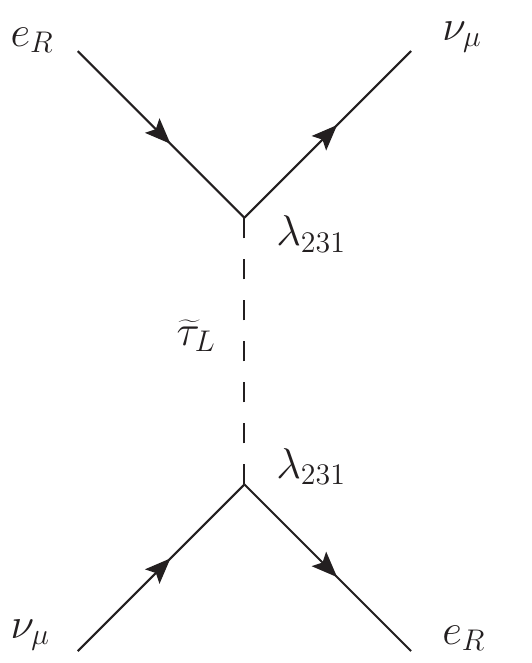}}
    \subfigure[$\ \lambda_{123},\lambda_{133}\neq 0$]{\includegraphics[width=3.5cm]{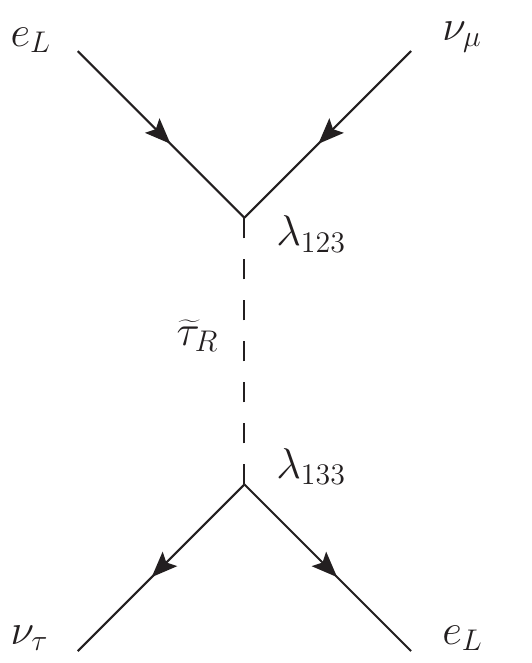}}
    \caption{NSI induced by charged-slepton exchange in the RPV3 framework for the cases considered in this work, where only the relevant $LLE$ couplings are taken to be nonzero.}
    \label{fig:LLE NSI}
\end{figure*}

\section{NSI Parameters} \label{sec:parameters}

With the relevant RPV3 cases defined in the previous section, we now turn to the corresponding NSI parameters. In cases (a)–(d), only diagonal NSI entries are generated, namely $\varepsilon_{\mu\mu}^{eL}$, $\varepsilon_{ee}^{eR}$, $\varepsilon_{\tau\tau}^{eL}$, and $\varepsilon_{\mu\mu}^{eR}$, respectively. Case (e) combines cases (a) and (c): it contains the diagonal contributions $\varepsilon_{\mu\mu}^{eL}$ from the coupling $\lambda_{123}$ and $\varepsilon_{\tau\tau}^{eL}$ from $\lambda_{133}$. In addition, the simultaneous presence of $\lambda_{123}$ and $\lambda_{133}$ induces an off-diagonal NSI parameter $\varepsilon_{\mu\tau}^{eL}$. Since the diagonal NSI parameters in case (e) coincide with those already present in case (a) and case (c), we will focus on the genuinely new contribution in case (e), namely $\varepsilon_{\mu\tau}^{eL}$.

Fig.~\ref{fig:LLE NSI} illustrates the NSI contributions in the various cases considered in this work. The corresponding effective four-fermion interactions and the resulting NSI parameters are given by
\begin{subequations}
\begin{align}
    {\cal L}_{\rm eff}^{\lambda_{123}} &=\frac{1}{2} \frac{|\lambda_{123}|^2}{m_{\widetilde{\tau}_{R}}^2} (\overline{e}_L \gamma^\mu e_L)(\overline{\nu}_{\mu L} \gamma_\mu \nu_{\mu L})~, \\
    {\cal L}_{\rm eff}^{\lambda_{131}}&=-\frac{1}{2} \frac{|\lambda_{131}|^2}{m_{\widetilde{\tau}_{L}}^2} (\overline{e}_R \gamma^\mu e_R)(\overline{\nu}_{e L} \gamma_\mu \nu_{e L})~, \\
    {\cal L}_{\rm eff}^{\lambda_{133}}&=\frac{1}{2} \frac{|\lambda_{133}|^2}{m_{\widetilde{\tau}_{R}}^2} (\overline{e}_L \gamma^\mu e_L)(\overline{\nu}_{\tau L} \gamma_\mu \nu_{\tau L})~, \\
    {\cal L}_{\rm eff}^{\lambda_{231}}&=-\frac{1}{2} \frac{|\lambda_{231}|^2}{m_{\widetilde{\tau}_{L}}^2} (\overline{e}_R \gamma^\mu e_R)(\overline{\nu}_{\mu L} \gamma_\mu \nu_{\mu L})~, \\
    {\cal L}_{\rm eff}^{\lambda_{123},\lambda_{133}}&=\frac{1}{2} \frac{|\lambda_{123}\lambda_{133}|}{m_{\widetilde{\tau}_{R}}^2} (\overline{e}_L \gamma^\mu e_L)(\overline{\nu}_{\tau L} \gamma_\mu \nu_{\mu L})~,
\label{eq:LLE four-fermion lagrangians}
\end{align}
\end{subequations}
and
\begin{subequations}
\begin{align}
    \varepsilon_{\mu\mu}^{\lambda_{123}}&=\varepsilon_{\mu\mu}^{eL,\lambda_{123}}=-\frac{1}{4\sqrt{2}G_F} \frac{|\lambda_{123}|^2}{m_{\widetilde{\tau}_{R}}^2}~, \\
    \varepsilon_{ee}^{\lambda_{131}}&=\varepsilon_{ee}^{eR,\lambda_{131}}=\frac{1}{4\sqrt{2}G_F} \frac{|\lambda_{131}|^2}{m_{\widetilde{\tau}_{L}}^2}~, \\
    \varepsilon_{\tau\tau}^{\lambda_{133}}&=\varepsilon_{\tau\tau}^{eL,\lambda_{133}}=-\frac{1}{4\sqrt{2}G_F} \frac{|\lambda_{133}|^2}{m_{\widetilde{\tau}_{R}}^2}~, \\
    \varepsilon_{\mu\mu}^{\lambda_{231}}&=\varepsilon_{\mu\mu}^{eR,\lambda_{231}}=\frac{1}{4\sqrt{2}G_F} \frac{|\lambda_{231}|^2}{m_{\widetilde{\tau}_{L}}^2}~, \\
    \varepsilon_{\mu\tau}^{\lambda_{123},\lambda_{133}}&=\varepsilon_{\mu\tau}^{eL,\lambda_{123},\lambda_{133}}=-\frac{1}{4\sqrt{2}G_F} \frac{|\lambda_{123}\lambda_{133}|}{m_{\widetilde{\tau}_{R}}^2}~.
\end{align}
\label{eq:LLE NSI parameters}
\end{subequations}

Here it is clear that, within the RPV3 framework, $\widetilde{\tau}_R$ exchange with coupling(s) $\lambda_{1\alpha 3}$ induces a negative NSI contribution to $\varepsilon^{eL}$, whereas $\widetilde{\tau}_L$ exchange with coupling(s) $\lambda_{\alpha 31}$ induces a positive contribution to $\varepsilon^{eR}$. Moreover, given the fact that $2|\lambda_{123}\lambda_{133}|\leq |\lambda_{123}|^2+|\lambda_{133}|^2$, we can infer that in case (e),
\begin{subequations}
\begin{align}
    \setcounter{equation}{4}
    |\varepsilon_{\mu\tau}^{\lambda_{123},\lambda_{133}}| &\leq \frac{1}{2} \left( |\varepsilon_{\mu\mu}^{\lambda_{123}}| + |\varepsilon_{\tau\tau}^{\lambda_{133}}| \right)~.
\end{align}
\label{eq:case (e) NSI limit}
\end{subequations}
This implies that, in case (e), the maximal off-diagonal NSI cannot exceed the average of the diagonal NSI contributions in cases (a) and (c). Once the additional constraints specific to case (e), which are absent in cases (a) and (c), such as $\tau\to\mu\gamma$ and $\tau\to \mu\nu\bar\nu$ are taken into account, the allowed upper bound on $\varepsilon_{\mu\tau}$ in case (e) can be further reduced (as discussed in subsequent sections).

\section{Constraints} \label{sec:constraints}

We now turn to the constraints relevant for the NSI scenarios discussed above. Since case (e) combines cases (a) and (c), any constraint applicable to case (a) or case (c) also applies to case (e). In the following, unless specifically stated otherwise, we will implicitly impose on case (e) all constraints that constrain case (a) and/or case (c).

\subsection{LHC constraints}
\label{sec:LHC constraints}

The LHC searches for sleptons in the $2\ell+\slashed E_T$ channel~\cite{ATLAS:2019lff,CMS:2020bfa}, based on slepton pair production, provide direct constraints on the charged-slepton masses. These limits are, however, model dependent and cannot be directly applied to the RPV3 spectra considered here. We therefore perform a recast of the ATLAS SR-SF-0J signal region of Ref.~\cite{ATLAS:2019lff}. The validation of the recast against the direct-slepton simplified model used in Ref.~\cite{ATLAS:2019lff}, together with the subsequent reinterpretation in the present RPV-$LLE$ scenarios, is described in Appendix~\ref{sec:validation}.

We then apply the same recast procedure to the RPV3 cases considered in this work and obtain exclusion contours in the $(m_{\widetilde{\tau}},\lambda)$ plane. The resulting 95\% CL lower bound is approximately
$m_{\widetilde{\tau}}\gtrsim 350$--$550~\mathrm{GeV}$, depending on the coupling assignment and on the relevant $\widetilde{\tau}_{R,L}$ decay channels. This bound is weaker than the $m_{\widetilde{\tau}}\sim 600$--$700~\mathrm{GeV}$ limits quoted in Refs.~\cite{ATLAS:2019lff,CMS:2020bfa}, because in the RPV3 setup the first two generations of sleptons are decoupled and do not contribute to the signal topologies. Nevertheless, the resulting LHC constraint remains at the sub-TeV level. The corresponding excluded regions are shown by the black-shaded regions in Fig.~\ref{fig:Results}.

\subsection{LEP contact interaction}

When the center-of-mass energy at an $e^+e^-$ collider exceeds the $Z$-boson mass, precision measurements of $e^+e^-\to f\bar f$ place stringent limits on possible four-fermion contact interactions~\cite{thelepcollaboration2004combination}. A standard parametrization of such contact interactions for $e^+e^-\to f\bar f$ is given by the effective Lagrangian~\cite{PhysRevLett.50.811}
\begin{equation}
    \mathcal{L}_{\text {eff }}=\frac{g^{2}}{\Lambda^{2}\left(1+\delta_{e f}\right)} \sum_{i, j=L, R} \eta_{i j}^{f}\left(\bar{e}_{i} \gamma^{\mu} e_{i}\right)\left(\bar{f}_{j} \gamma_{\mu} f_{j}\right)~,
\label{eq:LEP contact interaction Lagrangian}
\end{equation}
where $\delta_{ef}$ is the Kronecker delta, $g$ denotes the effective coupling strength, $\Lambda$ is the associated newphysics scale, and $\eta^f_{ij}=\pm 1$ or $0$ specifies the chiral structure of the interaction.

\begin{figure}[htbp]
    \centering
    \setcounter{subfigure}{1}
    \subfigure[]{\includegraphics[width=2.8cm]{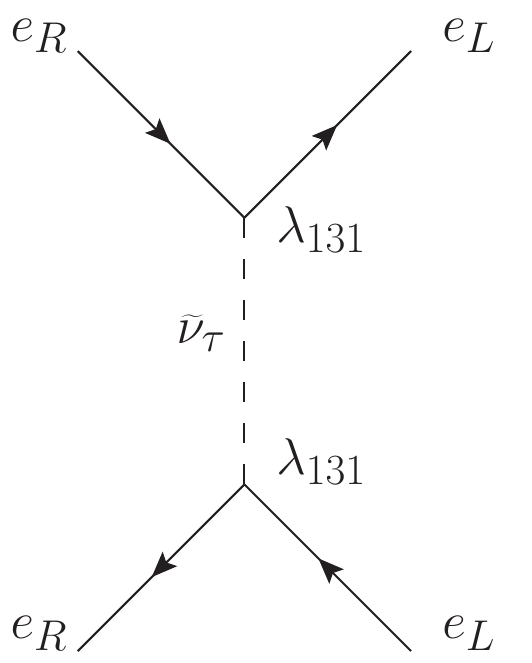}}
    \subfigure[]{\includegraphics[width=2.8cm]{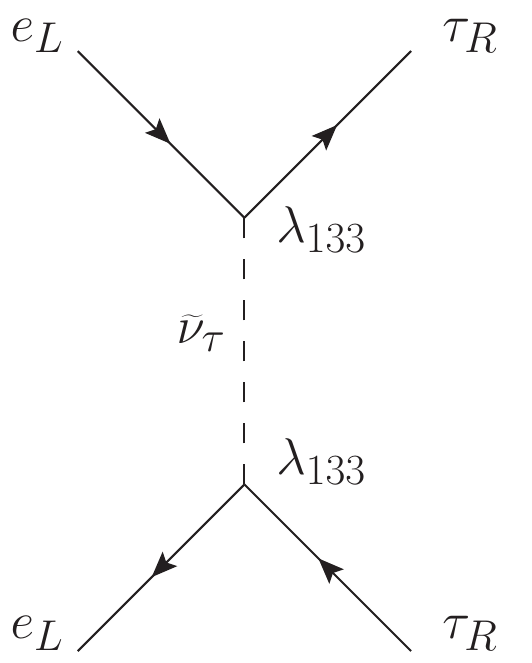}}
    \subfigure[]{\includegraphics[width=2.8cm]{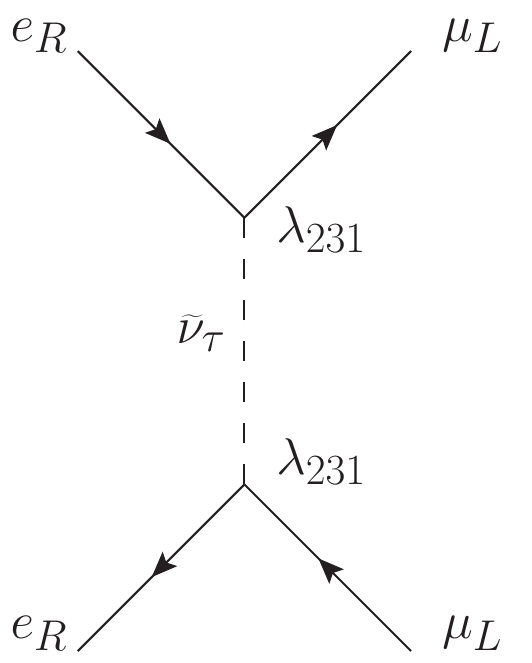}}
    \caption{Diagrams contributing to $e^+e^-\to f\bar f^\prime$ at LEP for the RPV3 cases considered in this work, with only the relevant $LLE$ couplings taken to be nonzero.}
    \label{fig:LEP contact interaction}
\end{figure}

LEP has set 95\% CL lower bounds on the contact-interaction scale $\Lambda$, adopting the conventional choice $g=\sqrt{4\pi}$~\cite{thelepcollaboration2004combination}. In the RPV3 framework, exchange of third-generation sfermions, such as $\widetilde{\nu}_\tau$ or $\widetilde{t}_L$, can contribute to $e^+e^-\to f\bar f^\prime$, where $f\bar f^\prime$ could be $d_i\bar d_j$ or $\ell_i\bar \ell_j$ ($i,j=1,2,3$). For the cases considered here, the relevant channels are $e^+e^-\to e^+e^-$, $e^+e^-\to \tau^+\tau^-$, and $e^+e^-\to \mu^+\mu^-$ for cases (b), (c), and (d), respectively; see Fig.~\ref{fig:LEP contact interaction}. No such process arises in case (a), since the required $\widetilde{\nu}_\tau$ interaction term is absent.

The effective Lagrangians for cases (b), (c), and (d) can be recast into the standard contact-interaction form after a Fierz rearrangement
\begin{subequations}
\begin{align}
    \setcounter{equation}{1}
    {\cal L}_{e^+ e^- \to e^+ e^-}^{\lambda_{131}}&=-\frac{|\lambda_{131}|^2}{2m_{\widetilde{\nu}_{\tau}}^2} (\bar{e}_L \gamma^\mu e_L) (\bar{e}_R \gamma_\mu e_R)~, \\
    {\cal L}_{e^+ e^- \to \tau^+ \tau^-}^{\lambda_{133}}&=-\frac{|\lambda_{133}|^2}{2m_{\widetilde{\nu}_{\tau}}^2} (\bar{e}_L \gamma^\mu e_L) (\bar{\tau}_R \gamma_\mu \tau_R)~, \\
    {\cal L}_{e^+ e^- \to \mu^+ \mu^-}^{\lambda_{231}}&=-\frac{|\lambda_{231}|^2}{2m_{\widetilde{\nu}_{\tau}}^2} (\bar{e}_R \gamma^\mu e_R) (\bar{\mu}_L \gamma_\mu \mu_L)~.
\end{align}
\label{eq:LEP contact interaction}
\end{subequations}
which can be directly compared with Eq.~\eqref{eq:LEP contact interaction Lagrangian}. We then obtain
\begin{subequations}
\begin{align}
    \setcounter{equation}{1}
    \frac{m_{\widetilde{\nu}_{\tau}}}{|\lambda_{131}|}&=\frac{\Lambda^-_{LR/RL}(e^+ e^- \to e^+ e^-)}{\sqrt{2}g}~, \\
    \frac{m_{\widetilde{\nu}_{\tau}}}{|\lambda_{133}|}&=\frac{\Lambda^-_{LR/RL}(e^+ e^- \to \tau^+ \tau^-)}{\sqrt{2}g}~, \\
    \frac{m_{\widetilde{\nu}_{\tau}}}{|\lambda_{231}|}&=\frac{\Lambda^-_{LR/RL}(e^+ e^- \to \mu^+ \mu^-)}{\sqrt{2}g}~,
\end{align}
\label{eq:LEP contact interaction constraints}
\end{subequations}
where $\Lambda^{-}$ denotes the contact-interaction scale for the choice $\eta^\ell_{LR}=\eta^\ell_{RL}=-1$. The LEP limits on $\Lambda$ (quoted for $g=\sqrt{4\pi}$) are given in Ref.~\cite{thelepcollaboration2004combination}. By matching to Eq.~\eqref{eq:LEP contact interaction constraints}, we translate these bounds into lower limits on $m_{\widetilde{\nu}_\tau}/|\lambda|$ for our cases (b), (c), and (d), summarized in Table~\ref{tab:LEP contact interaction constraints}.

{\renewcommand{\arraystretch}{1.8}
\begin{table}[htbp]
\caption{Constraints on $m_{\widetilde{\nu}_\tau}/|\lambda|$ inferred from LEP limits on $e^+e^-$ contact interactions.}
\begin{tabular}{c|c|c|c}
\hline
        Case & Process                     & LEP bound~\cite{thelepcollaboration2004combination}      & Constraint                                                    \\ \hline \hline
(b) & $e^+ e^- \to e^+ e^-$       & $\Lambda^-_{LR/RL}>10$ TeV  & $\frac{m_{\widetilde{\nu}_{\tau}}}{|\lambda_{131}|}>1.99$ TeV \\ \hline
(c) & $e^+ e^- \to \tau^+ \tau^-$ & $\Lambda^-_{LR/RL}>2.2$ TeV & $\frac{m_{\widetilde{\nu}_{\tau}}}{|\lambda_{133}|}>0.44$ TeV \\ \hline
(d) & $e^+ e^- \to \mu^+ \mu^-$   & $\Lambda^-_{LR/RL}>7.9$ TeV & $\frac{m_{\widetilde{\nu}_{\tau}}}{|\lambda_{231}|}>1.58$ TeV \\ \hline
\end{tabular}
\label{tab:LEP contact interaction constraints}
\end{table}
}

It should be emphasized that, within the RPV3 framework, the NSI parameters depend on the charged-slepton masses $m_{\widetilde{\tau}_R}$ or $m_{\widetilde{\tau}_L}$. The LEP contact-interaction bounds, on the other hand, constrain the sneutrino mass scale through limits on $m_{\widetilde{\nu}_\tau}/|\lambda|$. In Fig.~\ref{fig:Results}, we therefore show several representative choices of $m_{\widetilde{\nu}_\tau}$ for each case to illustrate how varying the sneutrino mass impacts the resulting NSI parameters. The corresponding LEP constraints are indicated by the blue dashed lines. We do not display these bounds as shaded exclusion regions, since their strength depends explicitly on $m_{\widetilde{\nu}_\tau}$ and they can be relaxed by taking $m_{\widetilde{\nu}_\tau}$ sufficiently large.

\subsection{$\ell_\alpha \to \ell_\beta \nu \bar{\nu}$}

\begin{figure}[htbp]
    \centering
    \subfigure[]{\includegraphics[width=2.75cm]{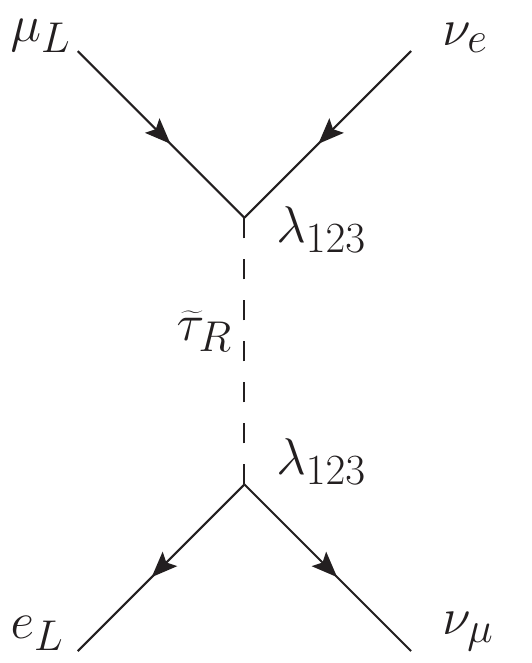}}
    \setcounter{subfigure}{2}
    \subfigure[]{\includegraphics[width=2.75cm]{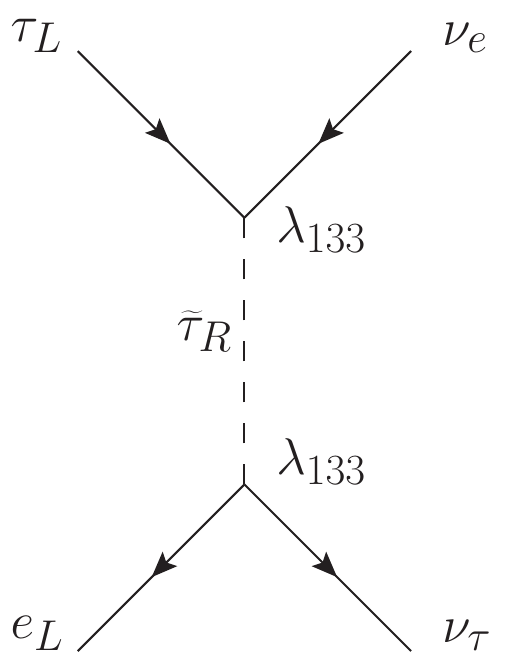}}
    \setcounter{subfigure}{4}
    \subfigure[]{\includegraphics[width=2.75cm]{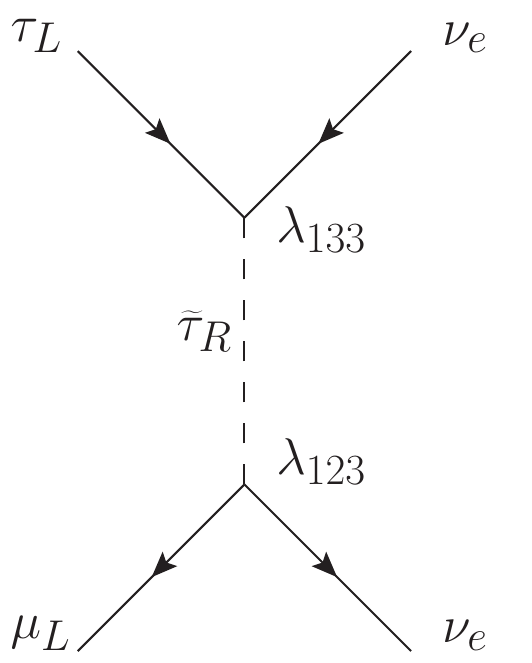}}
    \caption{Diagrams contributing to $\mu\to e\nu\bar\nu$, $\tau\to e\nu\bar\nu$, and $\tau\to \mu\nu\bar\nu$ mediated by $\widetilde{\tau}_R$ in the RPV3 setup considered in this work.}
    \label{fig:lvv}
\end{figure}

In this subsection, we discuss constraints from purely leptonic decays $\ell_\alpha\to \ell_\beta \nu\bar\nu$. Even when restricting to the $LLE$ interactions in the RPV3 framework, additional decay channels for $\tau$ and $\mu$ can arise, mediated by the right-handed stau $\widetilde{\tau}_R$. These contributions can modify the $\tau$ and $\mu$ lifetimes and affect lepton-flavor universality in leptonic $\tau$ decays, thereby providing constraints on the NSI parameters. As shown in Fig.~\ref{fig:lvv}, the relevant $LLE$ couplings in cases (a), (c), and (e) generate extra contributions to $\mu\to e\nu\bar\nu$, $\tau\to e\nu\bar\nu$, and $\tau\to \mu\nu\bar\nu$.

The effective four-fermion Lagrangians relevant for $\mu$ and $\tau$ leptonic decays in cases (a), (c), and (e) are given by
\begin{subequations}
\begin{align}
    {\cal L}^{\lambda_{123}}_{\mu \to e \nu \bar{\nu}}&=-\frac{|\lambda_{123}|^2}{2m^2_{\widetilde{\tau}_R}}(\overline{e}_L \gamma^\mu \nu_{e L})(\overline{\nu}_{\mu L} \gamma_\mu \mu_L)~, \\
    \setcounter{equation}{2}
    {\cal L}^{\lambda_{133}}_{\tau \to e \nu \bar{\nu}}&=-\frac{|\lambda_{133}|^2}{2m^2_{\widetilde{\tau}_R}}(\overline{e}_L \gamma^\mu \nu_{e L})(\overline{\nu}_{\tau L} \gamma_\mu \tau_L)~, \\
    \setcounter{equation}{4}
    {\cal L}^{\lambda_{123},\lambda_{133}}_{\tau \to \mu \nu \bar{\nu}}&=-\frac{|\lambda_{123} \lambda_{133}|}{2m^2_{\widetilde{\tau}_R}}(\overline{\mu}_L \gamma^\mu \nu_{e L})(\overline{\nu}_{e L} \gamma_\mu \tau_L)~,
\end{align}
\label{eq:lvv}
\end{subequations}
where we have rearranged the interactions into a form directly comparable to the standard vector-current parametrization of leptonic decays. Comparing with the SM Lagrangians for $\mu$ and $\tau$ decays,
\begin{subequations}
\begin{align}
    {\cal L}^{\rm SM}_{\mu \to e \nu \bar{\nu}}&=-2\sqrt{2} G_F (\overline{e}_L \gamma^\mu \nu_{e L})(\overline{\nu}_{\mu L} \gamma_\mu \mu_L)~, \\
    \setcounter{equation}{2}
    {\cal L}^{\rm SM}_{\tau \to e \nu \bar{\nu}}&=-2\sqrt{2} G_F (\overline{e}_L \gamma^\mu \nu_{e L})(\overline{\nu}_{\tau L} \gamma_\mu \tau_L)~, \\
    \setcounter{equation}{4}
    {\cal L}^{\rm SM}_{\tau \to \mu \nu \bar{\nu}}&=-2\sqrt{2} G_F (\overline{\mu}_L \gamma^\mu \nu_{\mu L})(\overline{\nu}_{\tau L} \gamma_\mu \tau_L)~,
\end{align}
\label{eq:lvv_SM}
\end{subequations}
we see that, for cases (a), (c), and (e), the coefficient $g_{LL}^V$ in the general four-fermion parametrization~\cite{Fetscher:1986uj,Kuno:1999jp} receives an additional contribution $\delta g_{LL}^V$,
\begin{subequations}
\begin{align}
    \delta g_{LL}^V (\mu \to e \nu \bar{\nu})&=\frac{1}{4\sqrt{2} G_F} \frac{|\lambda_{123}|^2}{m^2_{\widetilde{\tau}_R}}=-\varepsilon_{\mu\mu}^{\lambda_{123}}~, \\
    \setcounter{equation}{2}
    \delta g_{LL}^V (\tau \to e \nu \bar{\nu})&=\frac{1}{4\sqrt{2} G_F} \frac{|\lambda_{133}|^2}{m^2_{\widetilde{\tau}_R}}=-\varepsilon_{\tau\tau}^{\lambda_{133}}~, \\
    \setcounter{equation}{4}
    \delta g_{LL}^V (\tau \to \mu \nu \bar{\nu})&=\frac{1}{4\sqrt{2} G_F} \frac{|\lambda_{123} \lambda_{133}|}{m^2_{\widetilde{\tau}_R}}=-\varepsilon_{\mu\tau}^{\lambda_{123},\lambda_{133}}~.
\end{align}
\label{eq:delta g_LL^V in lvv}
\end{subequations}
However, the new contribution in our scenarios affects only the $g_{LL}^V$ coefficient. Owing to the standard normalization convention for the effective couplings~\cite{Fetscher:1986uj,Kuno:1999jp}, the Michel parameters remain unchanged in our setup. Nevertheless, case (c) is subject to a stringent constraint from $e$--$\mu$ universality in leptonic $\tau$ decays. The SM prediction, including mass effects, yields $\Gamma(\tau\to \mu\nu\bar\nu)/\Gamma(\tau\to e\nu\bar\nu)=97.26\%$~\cite{10.1093/ptep/ptac097}, while the current experiment prefers a slightly larger central value, $\Gamma(\tau\to \mu\nu\bar\nu)/\Gamma(\tau\to e\nu\bar\nu)=(97.62\pm 0.28)\%$~\cite{10.1093/ptep/ptac097}. In case (c), the partial width $\Gamma(\tau\to e\nu\bar\nu)$ is modified to $\Gamma(\tau\to e\nu\bar\nu)=(1+\delta g_{LL}^V(\tau\to e\nu\bar\nu))^2\, \Gamma^{\rm SM}(\tau\to e\nu\bar\nu)$. Allowing for a $3\sigma$ deviation, the $e$--$\mu$ universality test in $\tau$ decays implies
\begin{subequations}
\begin{align}
    \setcounter{equation}{2}
    |\varepsilon_{\tau\tau}^{\lambda_{133}}| &\leq 0.25\%~.
\end{align}
\label{lvv NSI constraints2}
\end{subequations}
Similarly, case (a) is constrained by the comparison between $\tau\to \mu\nu\bar\nu$ and $\mu\to e\nu\bar\nu$ decays~\cite{Barger:1989rk}, which is quantified by the ratio
\begin{align}
    R_{\tau / \mu} \equiv \frac{{\rm BR}(\tau \to \mu \nu \bar{\nu})_{\rm exp} / {\rm BR}(\tau \to \mu \nu \bar{\nu})_{\rm SM}}{{\rm BR}(\mu \to e \nu \bar{\nu})_{\rm exp} / {\rm BR}(\mu \to e \nu \bar{\nu})_{\rm SM}}~.
\end{align}
The current measurement yields a value of $R_{\tau/\mu}=1.0022\pm 0.0030$~\cite{Trifinopoulos:2018rna}. Following a similar derivation as that of Eq.~\eqref{lvv NSI constraints2}, and allowing for a $3\sigma$ deviation around the measured value, it translates into
\begin{subequations}
\begin{align}
    |\varepsilon_{\mu\mu}^{\lambda_{123}}| &\leq 0.56\%~.
\end{align}
\label{lvv NSI constraints3}
\end{subequations}

It should be noted that, unlike in the Zee model, where the new amplitude does not interfere with the SM one because of the chiral structure of the charged-lepton current~\cite{Babu:2019mfe}, the RPV3 contributions in cases (a) and (c) do interfere with the SM. As a result, the shifts in the $\mu$ and $\tau$ decay widths are of order $\delta g_{LL}^V$, rather than quadratic in $|\delta g_{LL}^V|^2$. For case (e), by contrast, the contribution to $\tau\to \mu\nu\bar\nu$ does not interfere with the SM amplitude due to the type of neutrinos involved. Furthermore, the constraints in Eq.~\eqref{lvv NSI constraints2} [derived for case (c)] and Eq.~\eqref{lvv NSI constraints3} [derived for case (a)] do not carry over to case (e), in contrast to the constraints discussed in the other subsections. In case (e), not only the widths $\Gamma(\tau\to e\nu\bar\nu)$ and $\Gamma(\mu\to e\nu\bar\nu)$ are modified, but $\Gamma(\tau\to \mu\nu\bar\nu)$ is modified to $(1+\delta g_{LL}^V(\tau \to \mu \nu \bar{\nu})^2) \, \Gamma^{\rm SM} (\tau \to \mu \nu \bar{\nu})$. Allowing for a $3\sigma$ deviation, the $e$--$\mu$ universality test in leptonic $\tau$ decays implies, for case (e),
\begin{subequations}
\begin{align}
    \setcounter{equation}{4}
    |\varepsilon_{\mu\tau}^{\lambda_{123},\lambda_{133}}| &\leq \sqrt{1.0123 \times (1 - \varepsilon_{\tau \tau}^{\lambda_{133}})^2 - 1}~,
\end{align}
\label{lvv NSI constraints4}
\end{subequations}
and the measurement of $R_{\tau / \mu}$ gives
\begin{subequations}
\begin{align}
    \setcounter{equation}{4}
    |\varepsilon_{\mu\tau}^{\lambda_{123},\lambda_{133}}| &\leq \sqrt{1.0112 \times (1 - \varepsilon_{\mu \mu}^{\lambda_{123}})^2 - 1}~.
\end{align}
\label{lvv NSI constraints5}
\end{subequations}
Based on the assumptions adopted in this work, both $\varepsilon_{\tau\tau}^{\lambda_{133}}$ and $\varepsilon_{\mu\mu}^{\lambda_{123}}$ are nonpositive definite [see Eqs.~(\ref{eq:LLE NSI parameters}a) and~(\ref{eq:LLE NSI parameters}c)]. As a consequence, Eqs.~\eqref{lvv NSI constraints4} and~\eqref{lvv NSI constraints5} translate into two comparatively weak bounds on the off-diagonal NSI parameter $\varepsilon_{\mu\tau}^{\lambda_{123},\lambda_{133}}$,
\begin{subequations}
\begin{align}
    \setcounter{equation}{4}
    |\varepsilon_{\mu\tau}^{\lambda_{123},\lambda_{133}}| &\leq 11.1\%~,
\end{align}
\label{eq:case (e) constraint1}
\end{subequations}
from the $e$--$\mu$ universality test, and
\begin{subequations}
\begin{align}
    \setcounter{equation}{4}
    |\varepsilon_{\mu\tau}^{\lambda_{123},\lambda_{133}}| &\leq 10.6\%~,
\end{align}
\label{eq:case (e) constraint2}
\end{subequations}
from $R_{\tau/\mu}$. These comparatively weak bounds are not unexpected, as they originate from ratios of decay rates, and case (e) happens to enhance both of them, albeit with different orders of magnitude. In Fig.~\ref{fig:Results}, the constraints from $R_{\tau/\mu}$ are shown by the orange-shaded regions, while the bound from $e$--$\mu$ universality is indicated by the yellow-shaded regions.

\subsection{$\tau \to \mu \gamma$}

The diagram in Fig.~\ref{fig:lvv}(e), relevant for case (e), is closely connected to the loop-induced LFV decay $\tau\to \mu\gamma$. This connection does not arise in cases (a) and (c), since the neutrino flavor assignments in Figs.~\ref{fig:lvv}(a) and~\ref{fig:lvv}(c) differ and therefore do not lead to the same $\tau\to \mu\gamma$ amplitude.

\begin{figure}[htbp]
    \centering
    \setcounter{subfigure}{4}
    \subfigure[]{\includegraphics[width=7cm]{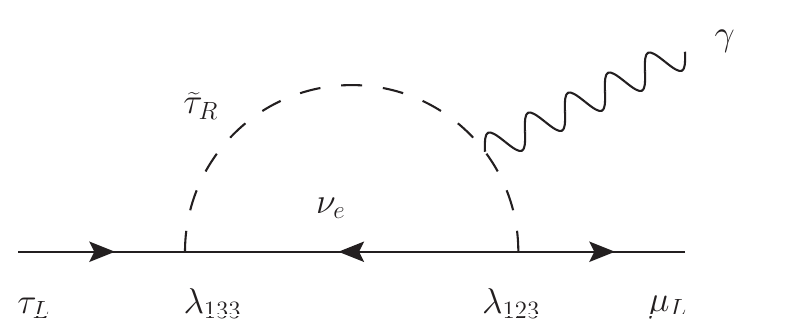}}
    \caption{Diagram contributing to the LFV decay $\tau\to\mu\gamma$ in the RPV3 scenario considered in this work.}
    \label{fig:tau_mugamma}
\end{figure}

The general RPV3 contribution to $\tau\to \mu\gamma$ has been derived in Ref.~\cite{Altmannshofer:2020axr}. Specifically, for our case (e), the corresponding partial width is
\begin{subequations}
\begin{align}
    \setcounter{equation}{4}
    \Gamma (\tau \to \mu \gamma) &\simeq \frac{\alpha m_\tau^5 G_F^2}{256 \pi^4} \frac{m_W^4}{g^4}\left| \frac{\lambda_{123} \lambda_{133}}{3 m_{\widetilde{\tau}_R}^2} \right|^2 \nonumber\\
    &= \frac{\alpha m_\tau^5 G_F^4}{72 \pi^4} \frac{m_W^4}{g^4} \varepsilon_{\mu \tau}^2~,
\end{align}
\label{eq:tau_mugamma}
\end{subequations}
where $\alpha$ and $g$ denote the electromagnetic and weak gauge couplings, respectively. Applying the current experimental data from the Belle Collaboration where ${\rm BR}(\tau \to \mu \gamma) < 4.2 \times 10^{-8}$~\cite{Belle:2021ysv}, Eq.~\eqref{eq:tau_mugamma} translates into the constraint
\begin{subequations}
\begin{align}
    \setcounter{equation}{4}
    |\varepsilon_{\mu\tau}^{\lambda_{123},\lambda_{133}}| &\leq 3.57\%~.
\end{align}
\label{eq:tau_mugamma case e}
\end{subequations}
The corresponding excluded region from $\tau\to \mu\gamma$ is indicated by the red-shaded region in Fig.~\ref{fig:Results}.

\subsection{Neutrino Mass}

\begin{figure}[htbp]
    \centering
    \includegraphics[width=8cm]{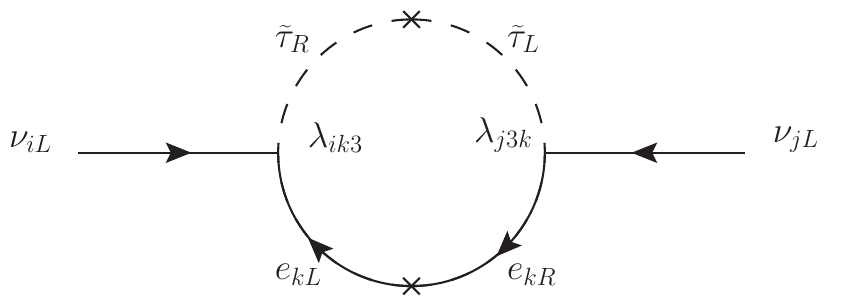}
    \caption{One-loop neutrino mass contribution induced by the $LLE$ interactions in the RPV3 framework.}
    \label{fig:LLE neutrino mass}
\end{figure}

The trilinear $LLE$ couplings in Eq.~\eqref{eq:RPVLLE} generate neutrino masses in the RPV3 framework at the one-loop level via lepton–slepton loops, as illustrated in Fig.~\ref{fig:LLE neutrino mass}. Starting from the general expressions in Refs.~\cite{Hall:1983id,Babu:1989px,Barbier:2004ez} and restricting to third-generation sfermion exchange~\cite{Altmannshofer:2020axr}, the neutrino mass matrix can be written in the simplified form
\begin{equation}
    M_{ij}^{\nu} \simeq \frac{1}{8 \pi^{2}}\left(\frac{A^{\tau}-\mu \tan \beta}{\overline{m}_{\widetilde{\tau}}^{2}}\right) \sum_{k} \lambda_{i k 3} \lambda_{j 3 k} m_{e_{k}} m_{\tau}~,
\label{eq:LLE neutrino mass}
\end{equation}
where $\overline{m}_{\widetilde{\tau}}$ denotes an average stau mass. In addition, we assume that the relevant soft trilinear $A$-terms~\cite{Barbier:2004ez} are proportional to the charged-lepton Yukawa couplings, working in the basis where the charged-lepton mass matrix is diagonal.

Within our setup, only the coupling $\lambda_{133}$ contributes to Eq.~\eqref{eq:LLE neutrino mass}. Imposing the cosmological limit on the sum of neutrino masses, $\sum_i m_{\nu_i}\lesssim 0.1~\mathrm{eV}$~\cite{Aghanim:2018eyx}, we obtain a constraint that applies only to cases (c) and (e),
\begin{subequations}
\begin{align}
    \setcounter{equation}{2}
    A^\tau - \mu \tan \beta \lesssim \mathcal{O}(\rm{GeV})~,
\end{align}
\label{eq:LLE neutrino mass constraint}
\end{subequations}
where we have assumed a representative value $\lambda_{133}/\overline{m}_{\widetilde{\tau}}\simeq 0.1/(1~\mathrm{TeV})$. The remaining cases do not contribute to Eq.~\eqref{eq:LLE neutrino mass} and are therefore unconstrained by neutrino mass considerations in our framework.

\subsection{Monophoton process}

In the RPV3 scenarios considered here, NSI involving electrons are directly associated with new contributions to the monophoton process $e^+e^-\to \nu\bar\nu\gamma$. The relevant diagrams are analogous to those in Fig.~\ref{fig:LLE NSI}, with a photon emitted from an external electron line or from the internal charged-slepton line. In the SM, $e^+e^-\to \nu\bar\nu\gamma$ proceeds through $s$-channel $Z$-boson exchange or $t$-channel $W$-boson exchange with a photon emitted from an electron leg or the intermediate $W$-boson. As a result, LEP monophoton measurements translate into constraints on the RPV3 contributions~\cite{Berezhiani:2001rs}. In cases (a), (c), (d), and (e), the final-state neutrinos changed flavors; in cases (b) and (d), the contribution involves a right-handed electron current. These characteristics ensure that the new RPV3 contributions do not interfere with the SM $W$-mediated process. Interference with the SM $Z$-exchange contribution is, in principle, allowed for flavor-diagonal final states; however, in the LEP monophoton kinematics, the dominant SM contribution often arises from radiative return to the $Z$ peak, so the impact of interference terms on the inclusive limits is numerically subleading.

The inclusive monophoton cross section can be decomposed as $\sigma=\sigma_{\rm SM}+\sigma_{\rm NS}$, where $\sigma_{\rm SM}$ denotes the SM prediction, and $\sigma_{\rm NS}$ collects the pure nonstandard contribution and possible interference with the SM amplitude. Requiring consistency with data, we impose
$|\sigma-\sigma_{\rm exp}|\leq \delta\sigma_{\rm exp}$,
where $\sigma_{\rm exp}\pm \delta\sigma_{\rm exp}$ is the experimentally measured cross section with its quoted uncertainty. Different from the Zee model~\cite{Babu:2019mfe}, where the light charged scalar mass is assumed around $100$--$500~\mathrm{GeV}$ and the heavier charged scalar is further constrained by LEP contact-interaction limits, the mediators relevant for our monophoton process are charged sleptons, typically with masses $m_{\widetilde{\tau}}\gtrsim \mathcal{O}(1~\mathrm{TeV})$. For such heavy mediators, even taking $\lambda\sim\mathcal{O}(1)$, we find that the resulting RPV3 contribution are about two order of magnitude smaller than the L3 experimental error~\cite{L3Collaboration1998,L3Collaboration2004}. Concretely, for the center of mass energy $\sqrt{s}=200~\mathrm{GeV}$, the mass of the charged slepton $m_{\widetilde{\tau}}=1~\mathrm{TeV}$, and the corresponding coupling $\lambda=1$, we obtain that the $\sigma_{\rm NS}$ is only $2\%$--$6\%$ of the $\delta \sigma_{\rm exp}$ across the five relevant cases. Therefore, within the parameter range displayed in Fig.~\ref{fig:Results}, the monophoton constraints are too weak and lie beyond the plotted region.

\subsection{Direct experimental searches}

Current neutrino data already place largely model-independent constraints on NSI, from a variety of experiments including IceCube~\cite{IceCube:2022pbe}, CHARM~II~\cite{Barranco:2007ej}, TEXONO~\cite{TEXONO:2010tnr}, BOREXINO~\cite{Borexino:2019mhy}, MINOS~\cite{MINOS:2013hmj}, and KamLAND~\cite{KamLAND:2008dgz,Bolanos:2008km,Farzan:2017xzy}, as well as the global-fit analyses~\cite{Coloma:2023ixt}. In particular, the global fit yields $-2\% \leq \varepsilon_{\mu\tau}^e \leq 1.2\%$, which directly constrains case (e). The bound $-1.5\% \leq \varepsilon_{\tau\tau}^e-\varepsilon_{\mu\mu}^e \leq 4.8\%$ translates into $\varepsilon_{\mu\mu}^e \geq -4.8\%$ for case (a), $\varepsilon_{\tau\tau}^e \geq -1.5\%$ for case (c), and $\varepsilon_{\mu\mu}^e \leq 1.5\%$ for case (d). For case (e), the constraint on $\varepsilon_{\tau\tau}^e-\varepsilon_{\mu\mu}^e$ can be completely circumvented from the cancellation between $\varepsilon_{\mu\mu}^e$ and $\varepsilon_{\tau\tau}^e$ as long as $|\lambda_{123}|=|\lambda_{133}|$. Nevertheless, simultaneously nonzero $\lambda_{123}$ and $\lambda_{133}$ remain constrained by $\tau\to \mu\nu\bar\nu$ and $\tau\to \mu\gamma$ in case (e), which in turn prevents $\varepsilon_{\mu\tau}^e$ from becoming large. This situation differs from the Zee model~\cite{Babu:2019mfe}, where an analogous cancellation is not viable: achieving sizable $\varepsilon_{\mu\mu}$ and $\varepsilon_{\tau\tau}$ typically correlates with a sizable $\varepsilon_{\mu\tau}$ that is tightly constrained by $\tau\to ee\mu$. By contrast, in general RPV3 settings, such cancellations can occur due to the larger freedom in the RPV parameter space. Notably, recent DUNE sensitivity studies incorporating $\nu_\tau$ detection at the far detector have shown an improved reach for propagation NSI, in particular for the off-diagonal parameter $\varepsilon_{\mu\tau}$~\cite{Yu:2025cyx}. In Fig.~\ref{fig:Results}, we also include the projected $2\sigma$ sensitivity at DUNE for an exposure of $336~\mathrm{kt\cdot MW\cdot yr}$~\cite{DUNE:2020ypp,DUNE:2021cuw,DUNE:2021mtg,Chatterjee:2021wac}. Finally, note that in cases (a) and (c) the total NSI is governed by $\varepsilon_{\alpha\beta}^{eL}$, whereas in cases (b) and (d) it is governed by $\varepsilon_{\alpha\beta}^{eR}$; accordingly, for each case we apply the most relevant experimental bound. For the CHARM~II constraint on $\varepsilon_{\mu\mu}^{eR}$~\cite{Barranco:2007ej}, we adopt the updated range $-1.7\%<\varepsilon_{\mu\mu}^{eR}<3.8\%$~\cite{Babu:2019mfe}, obtained using the latest value of $s_w^2$.

In Fig.~\ref{fig:Results}, the IceCube bounds are shown as the brown-shaded regions. The CHARM~II, TEXONO, BOREXINO, MINOS, and KamLAND constraints are indicated by the magenta-, purple-, green-, pink-, and cyan-shaded regions, respectively. The global-fit limits are displayed by the violet-shaded regions, while the projected DUNE sensitivities are marked by the gray dashed lines.

\subsection{Anomalous magnetic moment of the electron and muon}

\begin{figure*}[htbp]
    \centering
    \subfigure[]{\includegraphics[width=4cm]{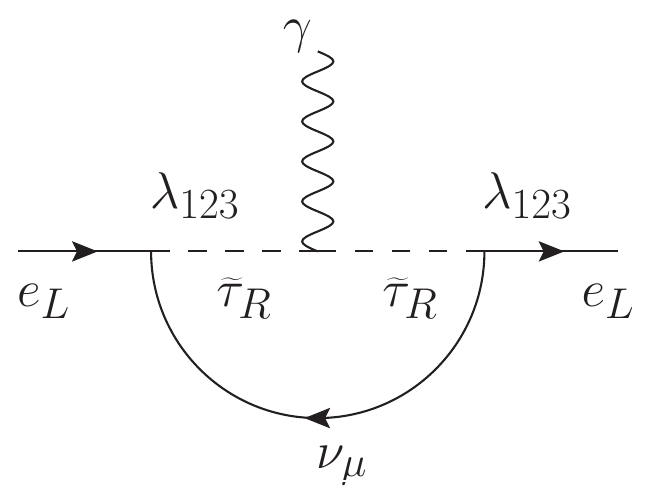}}
    \setcounter{subfigure}{0}\renewcommand{\thesubfigure}{(b\arabic{subfigure})}\makeatletter
    \subfigure[]{\includegraphics[width=4cm]{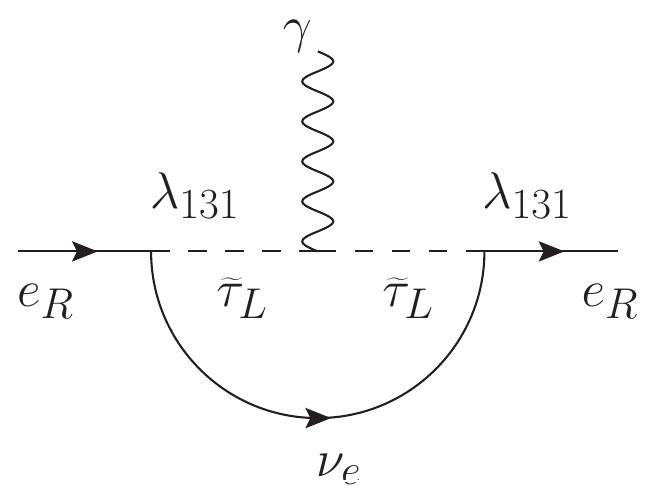}}
    \subfigure[]{\includegraphics[width=4cm]{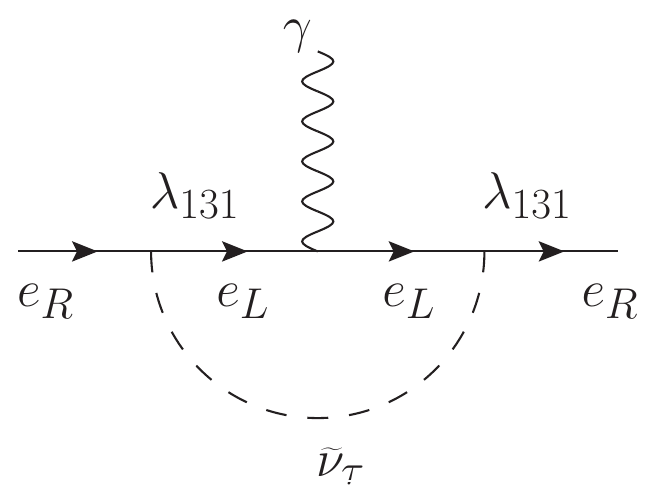}}
    \subfigure[]{\includegraphics[width=4cm]{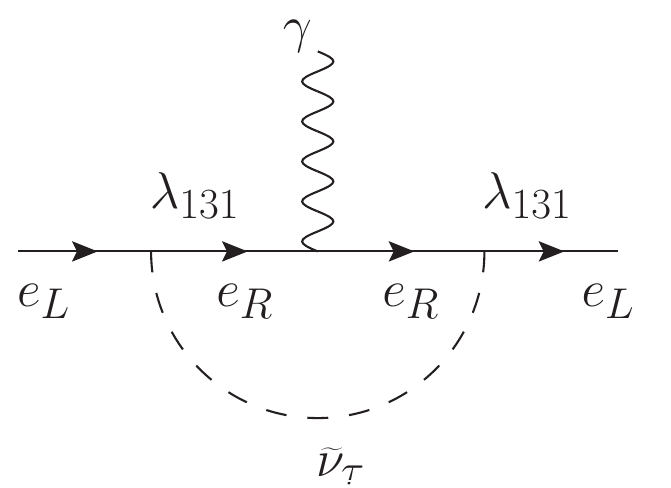}}
    \setcounter{subfigure}{0}\renewcommand{\thesubfigure}{(c\arabic{subfigure})}\makeatletter
    \subfigure[]{\includegraphics[width=4cm]{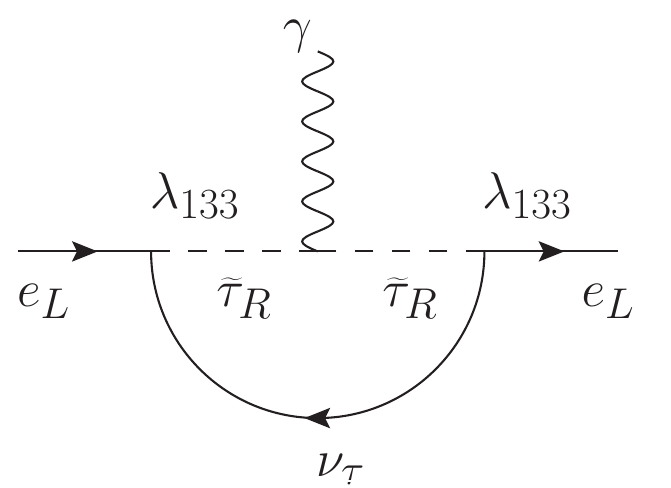}}
    \subfigure[]{\includegraphics[width=4cm]{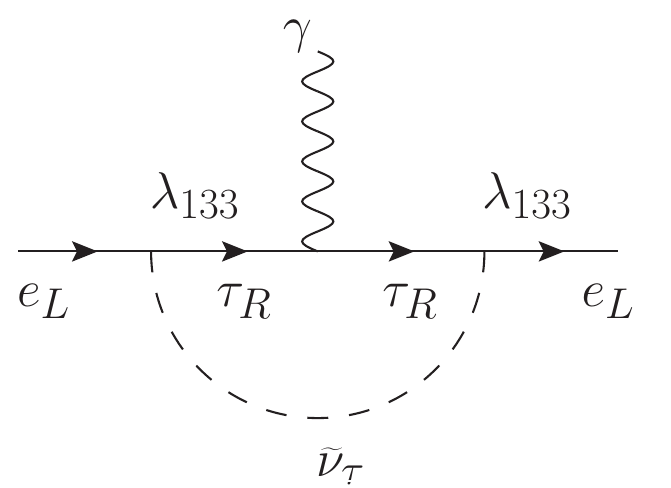}}
    \setcounter{subfigure}{0}\renewcommand{\thesubfigure}{(d\arabic{subfigure})}\makeatletter
    \subfigure[]{\includegraphics[width=4cm]{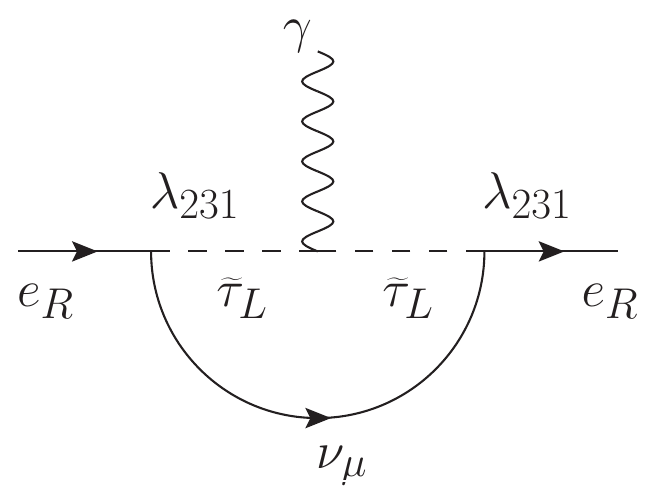}}
    \subfigure[]{\includegraphics[width=4cm]{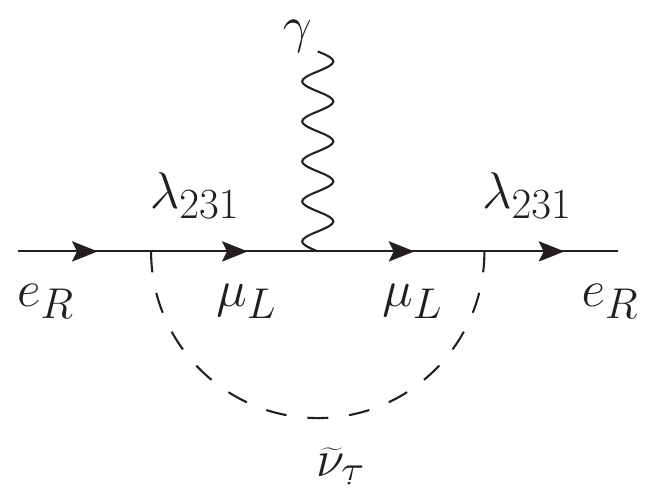}}
    \caption{Diagrams contributing to $(g-2)_e$ from $LLE$ interactions in the cases considered in this work.}
    \label{fig:electron g-2}
\end{figure*}

\begin{figure}[htbp]
    \centering
    \subfigure[]{\includegraphics[width=4cm]{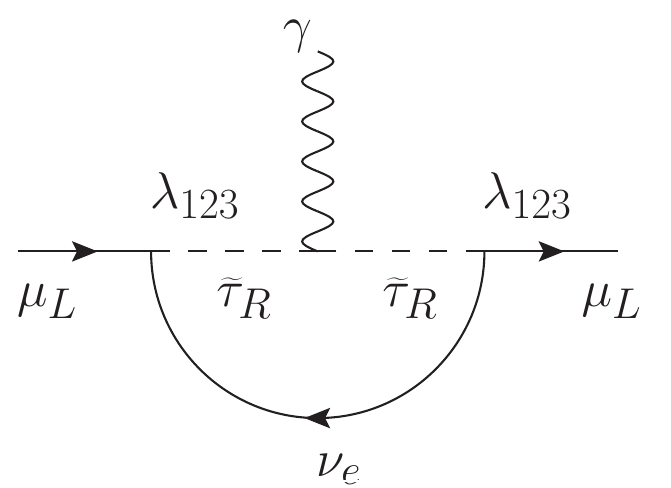}}
    \setcounter{subfigure}{3}
    \subfigure[]{\includegraphics[width=4cm]{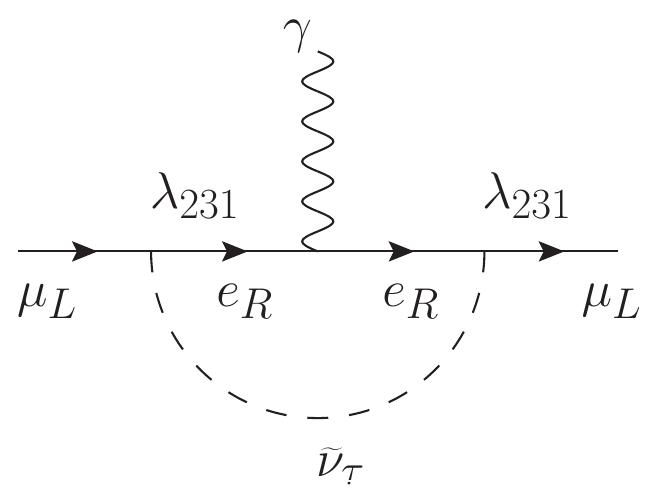}}
    \caption{Diagrams contributing to $(g-2)_\mu$ from $LLE$ interactions in the cases considered in this work.}
    \label{fig:muon g-2}
\end{figure}

In the RPV3 framework, contributions to the electron and muon anomalous magnetic moments, $(g-2)_e$ and $(g-2)_\mu$, can arise from both $\lambda$ and $\lambda^\prime$ interactions~\cite{Kim:2001se}. Restricting to the $LLE$ sector relevant for our analysis, the contributions to $(g-2)_e$ in the cases considered here are summarized in Fig.~\ref{fig:electron g-2} and can be written as
\begin{subequations}
\begin{align}
    \Delta a_e^{\lambda_{123}}&=\frac{m_e^2}{96\pi^2} \left( - \frac{|\lambda_{123}|^2}{m_{\widetilde{\tau}_R}^2} \right)~, \\
    \Delta a_e^{\lambda_{131}}&=\frac{m_e^2}{96\pi^2} \left( - \frac{|\lambda_{131}|^2}{m_{\widetilde{\tau}_L}^2} + \frac{4|\lambda_{131}|^2}{m_{\widetilde{\nu}_\tau}^2} \right)~, \\
    \Delta a_e^{\lambda_{133}}&=\frac{m_e^2}{96\pi^2} \left( - \frac{|\lambda_{133}|^2}{m_{\widetilde{\tau}_R}^2} + \frac{2|\lambda_{133}|^2}{m_{\widetilde{\nu}_\tau}^2} \right)~, \\
    \Delta a_e^{\lambda_{231}}&=\frac{m_e^2}{96\pi^2} \left( - \frac{|\lambda_{231}|^2}{m_{\widetilde{\tau}_L}^2} + \frac{2|\lambda_{231}|^2}{m_{\widetilde{\nu}_\tau}^2} \right)~.
\end{align}
\label{eq:electron g-2}
\end{subequations}
Likewise, the corresponding contributions to $(g-2)_\mu$ are shown in Fig.~\ref{fig:muon g-2} and are given by
\begin{subequations}
\begin{align}
    \Delta a_\mu^{\lambda_{123}}&=\frac{m_\mu^2}{96\pi^2} \left( - \frac{|\lambda_{123}|^2}{m_{\widetilde{\tau}_R}^2} \right)~, \\
    \setcounter{equation}{3}
    \Delta a_\mu^{\lambda_{231}}&=\frac{m_\mu^2}{96\pi^2} \left( \frac{2|\lambda_{231}|^2}{m_{\widetilde{\nu}_\tau}^2} \right)~.
\end{align}
\label{eq:muon g-2}
\end{subequations}
In this setup, only $\lambda_{123}$ [corresponding to case (a), and hence also to case (e)] and $\lambda_{231}$ [corresponding to case (d)] contribute to $(g-2)_\mu$.

Based on the discussion in the previous subsections, the neutrino NSI parameters in our setup are typically constrained to be at most at the few-percent level. Although the $m_{\widetilde{\nu}_\tau}$, which is not relevant for NSI, can enter the contributions to $(g-2)_e$ and $(g-2)_\mu$ in some cases, it is still instructive to estimate the stau-induced effect, which in the present context is directly tied to the NSI parameters. In particular, combining Eq.~\eqref{eq:LLE NSI parameters} with the first terms in Eq.~\eqref{eq:electron g-2} and Eq.~(\ref{eq:muon g-2}a), one finds
\begin{align}
    \Delta a_e = -\frac{m_e^2 G_F}{12\sqrt{2} \pi^2} |\varepsilon|~,
\label{eq:NSI g-2 e}
\end{align}
\begin{align}
    \Delta a_\mu = -\frac{m_\mu^2 G_F}{12\sqrt{2} \pi^2} |\varepsilon|~,
\label{eq:NSI g-2 mu}
\end{align}
where $\varepsilon$ denotes the NSI parameter in each relevant case. Eqs.~\eqref{eq:NSI g-2 e} and~\eqref{eq:NSI g-2 mu} show that the stau contribution (and thus, the NSI-linked effect in this setup) is negative for both the electron and the muon $(g-2)$. Furthermore, taking a representative value of the NSI parameter $|\varepsilon|=5\%$ and inserting the standard values of $m_e$, $m_\mu$, and $G_F$, we obtain $\Delta a_e \simeq -1\times 10^{-15}$ and $\Delta a_\mu \simeq -4\times 10^{-11}$. These estimates are well below the current experimental sensitivities: $\Delta a_e^{\rm Cs}=(-8.7\pm 3.6)\times 10^{-13}$~\cite{Parker2018MeasurementOT} based on Cs measurement, $\Delta a_e^{\rm Rb}=(4.8\pm 3.0)\times 10^{-13}$~\cite{morel2020determination} based on Rb measurement, and $\Delta a_\mu=(38\pm 63)\times 10^{-11}$ from the Fermilab muon $(g-2)$ measurement~\cite{Aliberti:2025beg,Muong-2:2025xyk}. Numerically, the contribution in Eq.~\eqref{eq:NSI g-2 e} amounts to at most $\sim 0.2\%$ of the central value of $\Delta a_e$ and $\sim 0.3\%$ of its $1\sigma$ uncertainty, while Eq.~\eqref{eq:NSI g-2 mu} gives at most $\sim 10.5\%$ of the central value of $\Delta a_\mu$ and $\sim 6.3\%$ of its $1\sigma$ uncertainty.

On the other hand, the sneutrino contributions in Eq.~\eqref{eq:electron g-2} and Eq.~\eqref{eq:muon g-2} are also subject to constraints from the LEP contact interaction experiment as shown in Table~\ref{tab:LEP contact interaction constraints}. Specifically, in case (b), the $\widetilde{\nu}_{\tau}$ contribution to $\Delta a_e$ is less than $2.8 \times 10^{-16}$; in case (c), the $\widetilde{\nu}_{\tau}$ contribution to $\Delta a_e$ is less than $2.8 \times 10^{-15}$; in case (d), the $\widetilde{\nu}_{\tau}$ contribution is less than $2.2 \times 10^{-16}$ for $\Delta a_e$ and less than $9.4 \times 10^{-12}$ for $\Delta a_\mu$. These values are only at the order of $0.01\%$--$1\%$ of the corresponding central values of $\Delta a_e$ and $\Delta a_\mu$.

As a result, the $(g-2)_e$ and $(g-2)_\mu$ sensitive regions in the $(m_{\widetilde{\tau}},\lambda)$ parameter space are significantly larger in scale compared to the NSI sensitive regions in the RPV3 framework. Hence, the constraints from $(g-2)_e$ and $(g-2)_\mu$ are not depicted in Fig.~\ref{fig:Results}. Besides, this feature reinforces the notion in Refs.~\cite{BhupalDev:2021ipu,Afik:2022vpm} that the primary source of the $(g-2)$ does not originate from the stau contribution, but rather from the sneutrino contribution\footnote{In Refs.~\cite{BhupalDev:2021ipu,Afik:2022vpm}, the constraints from the LEP contact interaction do not apply, as the scenarios therein are specifically focused on the $\mu \tau$ sector of the $\lambda$ couplings.} within the framework of RPV3.

\section{Numerical Results} \label{sec:numerical}

\begin{figure*}[htbp]
    \centering
    \subfigure[]{\includegraphics[width=5.8cm]{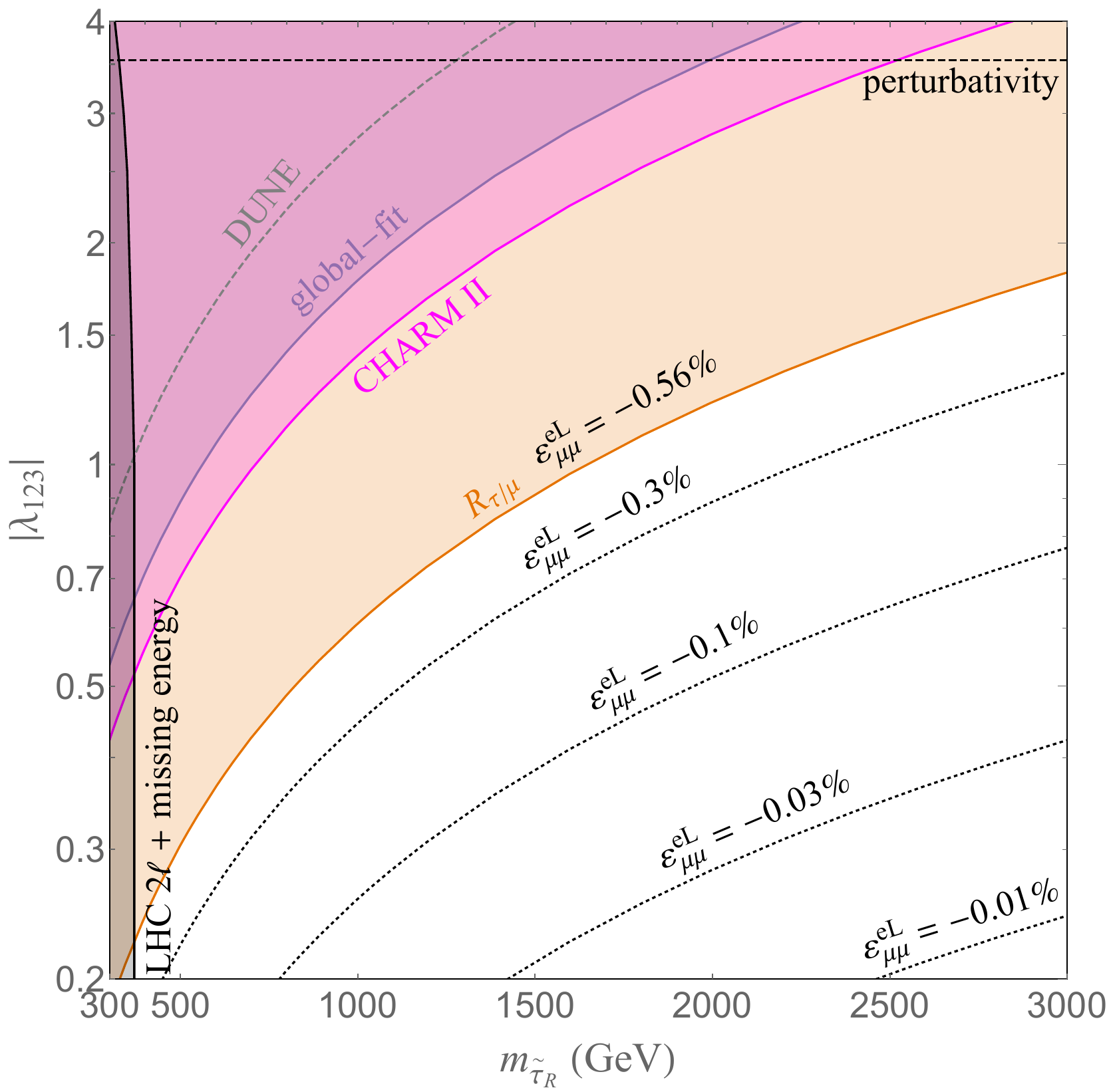}}
    \subfigure[]{\includegraphics[width=5.8cm]{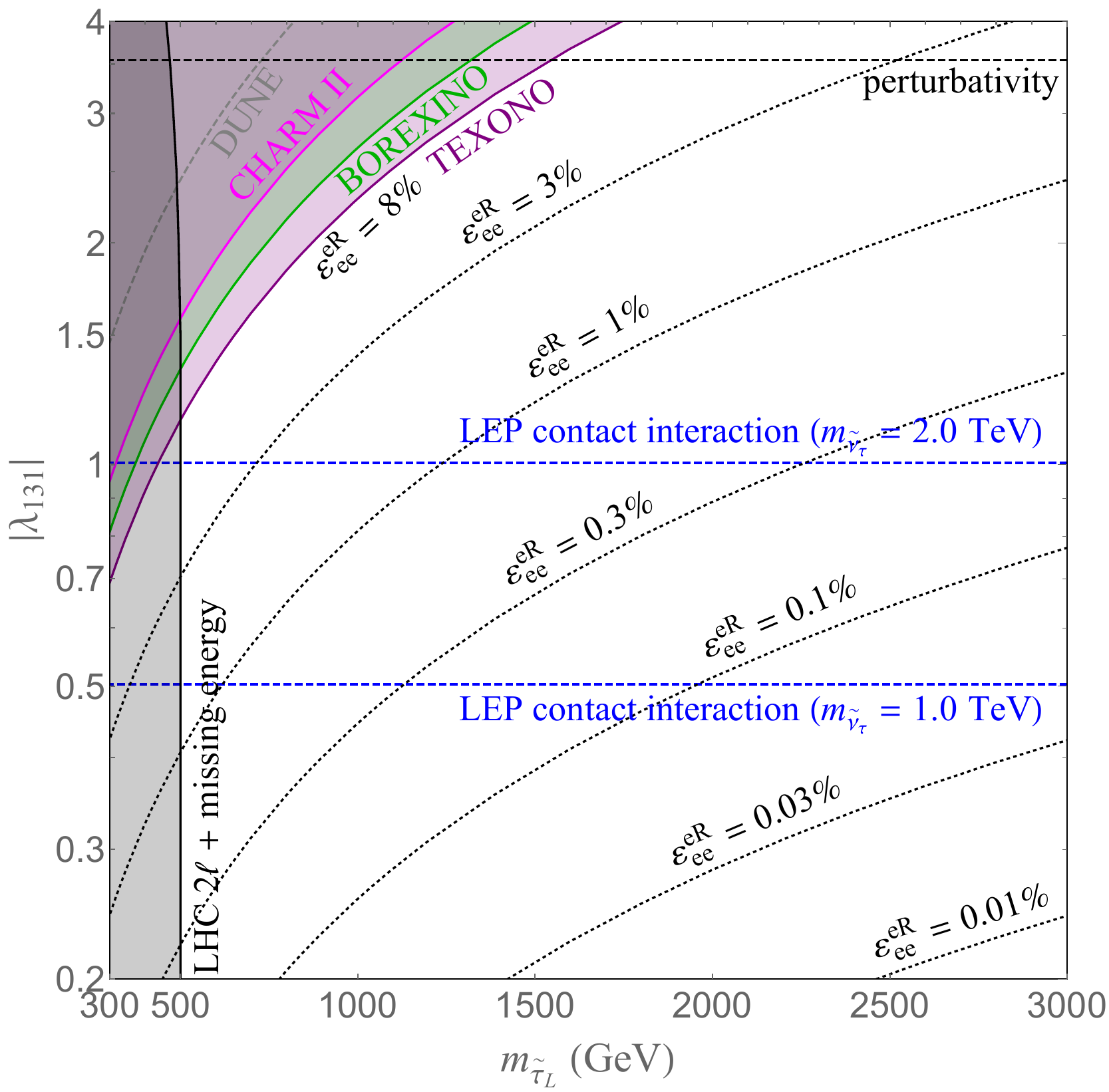}}
    \subfigure[]{\includegraphics[width=5.8cm]{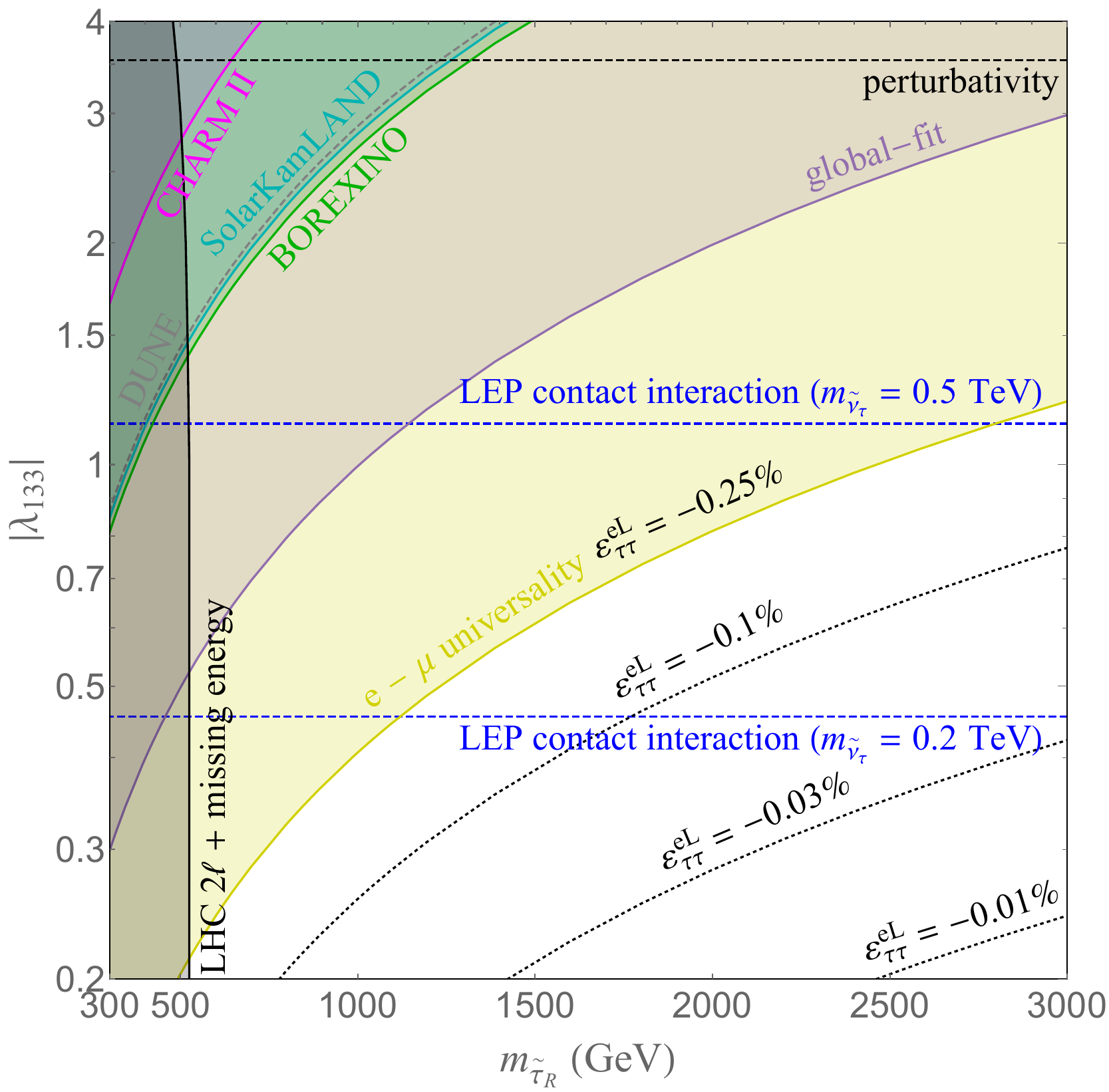}}
    \subfigure[]{\includegraphics[width=5.8cm]{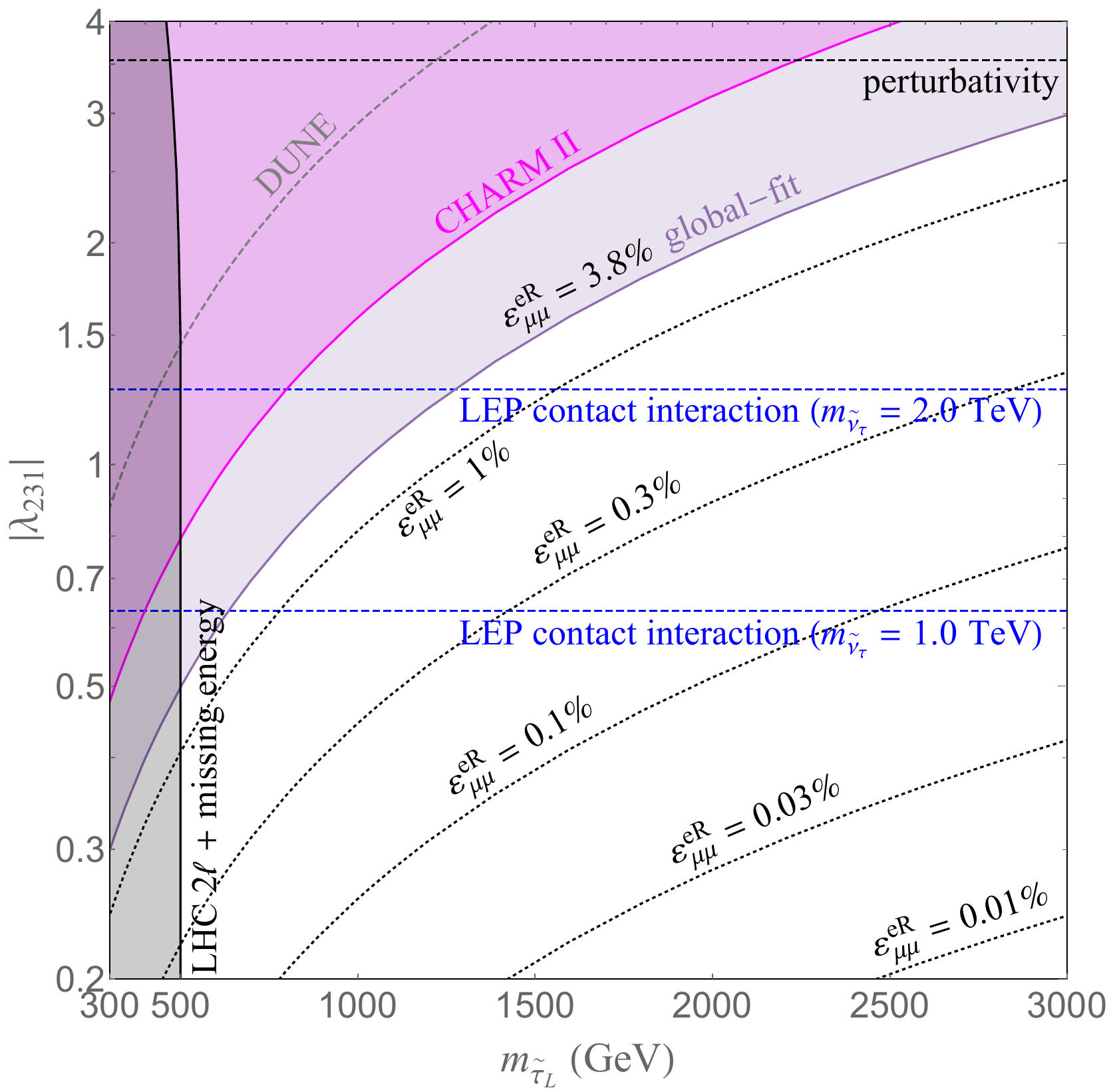}}
    \subfigure[]{\includegraphics[width=6cm]{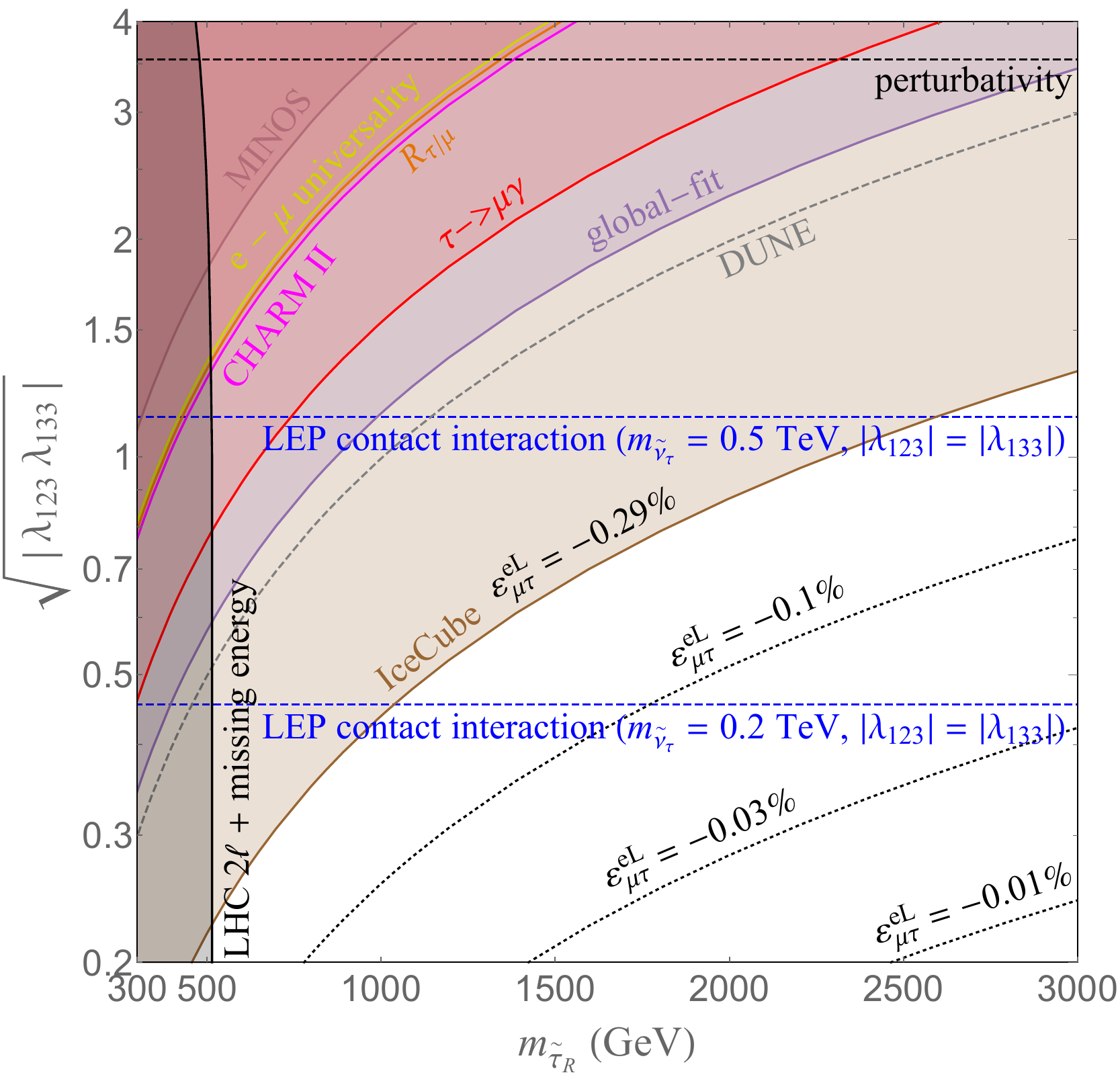}}
    \caption{Constraints and allowed regions for cases (a)–(e) in the corresponding $(m_{\widetilde{\tau}},\lambda)$ plane. The black dotted curves show representative values of the resulting NSI parameters. Exclusions from the LHC are indicated by the black-shaded regions, while LEP bounds are shown by the blue dashed lines. Constraints from $R_{\tau/\mu}$ and from $e$--$\mu$ universality are shown by the orange- and yellow-shaded regions, respectively. The bounds from IceCube, CHARM~II, TEXONO, BOREXINO, MINOS, and KamLAND correspond to the brown-, magenta-, purple-, green-, pink-, and cyan-shaded regions, respectively, and the global-fit limits are shown by the violet-shaded regions. The projected DUNE sensitivities are indicated by the gray dashed lines. The horizontal black dashed line denotes the perturbativity limit $\lambda=\sqrt{4\pi}$.}
    \label{fig:Results}
\end{figure*}

In Fig.~\ref{fig:Results}, we display the constraints and allowed regions for cases (a)–(e) in the corresponding $(m_{\widetilde{\tau}},\lambda)$ parameter space. The black dotted curves indicate illustrative values of the resulting NSI parameters. The LHC exclusions are shown as the black-shaded regions, while the LEP bounds are marked by the blue dashed lines. Constraints from $R_{\tau/\mu}$ and from $e$--$\mu$ universality are indicated by the orange- and yellow-shaded regions, respectively. The bounds from IceCube, CHARM~II, TEXONO, BOREXINO, MINOS, and KamLAND are shown by the brown-, magenta-, purple-, green-, pink-, and cyan-shaded regions, respectively, and the global-fit constraints are displayed by the violet-shaded regions. The projected DUNE sensitivities are indicated by the gray dashed lines. Finally, the horizontal black dashed line denotes the perturbativity limit $\lambda=\sqrt{4\pi}$.

Within our working setup, which is characterized by an RPV3-like spectrum and NSI predominantly induced by $LLE$ interactions, we find that the largest magnitudes of the NSI parameters consistent with current constraints are

\begin{itemize}
\item In case (a) where $\lambda_{123}\neq 0$: $\varepsilon_{\mu\mu}=-0.56\%$, dominated by the constraint from the ratio of $\tau\to \mu\nu\bar\nu$ to $\mu\to e\nu\bar\nu$;
\item In case (b) where $\lambda_{131}\neq 0$: $\varepsilon_{ee}=8\%$, set primarily by the TEXONO bound;
\item In case (c) where $\lambda_{133}\neq 0$: $\varepsilon_{\tau\tau}=-0.25\%$, limited by $e$--$\mu$ universality in $\tau$ decays;
\item In case (d) where $\lambda_{231}\neq 0$: $\varepsilon_{\mu\mu}=3.8\%$, driven by the global-fit constraint;
\item In case (e) where $\lambda_{123}\lambda_{133}\neq 0$: $\varepsilon_{\mu\tau}=-0.29\%$, dominated by the IceCube limit.
\end{itemize}

In the scenarios considered here, cases (b) with $\lambda_{131}\neq 0$ and (d) with $\lambda_{231}\neq 0$ receive their NSI contribution from $\widetilde{\tau}_L$, leading to a positive $\varepsilon^{eR}$. By contrast, in case (a) with $\lambda_{123}\neq 0$, case (c) with $\lambda_{133}\neq 0$, and case (e) with $\lambda_{123},\lambda_{133}\neq 0$, the NSI arises from $\widetilde{\tau}_R$, resulting in a negative $\varepsilon^{eL}$. Numerically, the allowed NSI magnitudes in cases (b) and (d) (with NSI induced by $\widetilde{\tau}_L$ exchange) are relatively larger than those in cases (a), (c), and (e) (with NSI induced by $\widetilde{\tau}_R$ exchange). This difference is mainly driven by the additional charged-lepton decay constraints that accompany the $\widetilde{\tau}_R$-mediated scenarios, which more tightly limit the corresponding parameter space and prevent $|\varepsilon|$ from reaching comparable values. Although the $\widetilde{\tau}_L$- and $\widetilde{\tau}_R$-mediated interactions arising from the expansion of the $LLE$ term appear rather similar, as both contain couplings of a scalar to a charged lepton and a neutrino, the allowed NSI effects can differ substantially between them. To better understand this pattern, we now examine the underlying interaction structure at the level of the Lagrangian and symmetry:
\\
\\
{\bf (i) Correlated vertices as the origin of the different constraints}

In particular, in the $LLE$ interaction Lagrangian, a given coupling $\lambda_{ijk}$ always generates a correlated set of component vertices. This is, in fact, a characteristic feature of RPV interactions: several interaction terms in the expanded Lagrangian often share the same coupling, rather than being controlled by independent coupling parameters.

For the $\widetilde{\tau}_R$-mediated cases (with $k=3$), the relevant NSI vertex involving $\widetilde{\tau}_R^\ast,\bar{\nu}_{iL}^{c},e_{jL}$ is unavoidably accompanied, by the $(i\leftrightarrow j)$ antisymmetry, by the partner vertex with the lepton and neutrino flavors exchanged, $\widetilde{\tau}_R^\ast,\bar{\nu}_{jL}^{c},e_{iL}$. As a result, once the coupling responsible for NSI is switched on, one typically also induces additional tree-level charged-lepton processes (e.g. $\ell_j\to \ell_i\nu\bar{\nu}$ or related universality ratios) mediated by the same $\widetilde{\tau}_R$, which leads to stronger bounds on $|\varepsilon|$ in the $\widetilde{\tau}_R$-dominated cases.

By contrast, for the $\widetilde{\tau}_L$-mediated cases the NSI-relevant vertices involve $\widetilde{\tau}_L$ carrying the $i$ or $j$ index (rather than the $k$ index), so the accompanying $(i\leftrightarrow j)$ term does not generate the same pattern of correlated charged-lepton decays, allowing comparatively larger NSI magnitudes.
\\
\\
{\bf (ii) The superpotential origin of the different correlated vertices}

Although the $\widetilde{\tau}_L$- and $\widetilde{\tau}_R$-mediated interactions can both generate scalar--lepton--neutrino couplings and hence similar NSI structures at low energy, the accompanying interaction vertices and the resulting constraints are different. The origin of this difference can be understood by tracing the relevant vertices back to the superpotential term $W\supset \tfrac12\lambda_{ijk}\hat L_i\hat L_j\hat E_k^c$. Here $\hat L_i$ is an $SU(2)_L$ doublet superfield, whereas $\hat E_k^c$ is an $SU(2)_L$ singlet. Consequently, the left-handed stau $\widetilde{\tau}_L$ is part of the same electroweak doublet as the tau sneutrino, $(\widetilde{\nu}_\tau,\widetilde{\tau}_L)$, mirroring the SM lepton doublet structure, while $\widetilde{\tau}_R$ originates from the singlet superfield $\widetilde{E}^c$ and therefore does not come with a sneutrino partner.

Although the component interactions mediated by $\widetilde{\tau}_L$ and $\widetilde{\tau}_R$ can look superficially similar, the antisymmetric $SU(2)_L$ contraction in $\hat L_i\hat L_j$ implies that switching on a given $\lambda_{ijk}$ necessarily activates a correlated set of doublet-related vertices, with the $(i\leftrightarrow j)$ piece fixed by gauge structure. For $\widetilde{\tau}_L$ this correlation is manifest already at the sfermion level through its doublet partner $(\widetilde{\nu}_i,\widetilde{e}_{jL})$, whereas for $\widetilde{\tau}_R$ (from the singlet superfield $\hat E^c$), the correlation instead reflects the SM doublet structure carried by the $\hat L$ fields (so that the antisymmetry under $i\leftrightarrow j$ effectively induces the exchange of the lepton and neutrino flavors). In practice, this correlation is precisely what ties the NSI to additional charged-lepton processes more tightly in the $\widetilde{\tau}_R$-mediated cases than in the $\widetilde{\tau}_L$-mediated ones.
\\
\\
{\bf (iii) Mediator embedded in an $SU(2)_L$ doublet versus external singlet scalar}

For cases (b) and (d), where $\widetilde{\tau}_L$ mediates the NSI, our results resemble those found in the Zee model~\cite{Babu:2019mfe}. In both cases, $\varepsilon_{e e}$ can reach values $\sim 8\%$, and $\varepsilon_{\mu \mu}$ can reach values $\sim 3.8\%$. In contrast, $\varepsilon_{\tau\tau}$ is significantly more limited in the RPV3 setup. Although case (c) allows a $\widetilde{\tau}_R$ contribution to $\varepsilon_{\tau\tau}$, it is strongly constrained by $e$--$\mu$ universality and cannot approach the $\varepsilon_{\tau\tau}\simeq 9.3\%$ values attainable in the Zee model~\cite{Babu:2019mfe}. The limitation of $\varepsilon_{\tau\tau}$ stems from the fact that an analogous $\widetilde{\tau}_L$ contribution to $\varepsilon_{\tau\tau}$ is absent here due to the antisymmetric property of the $\lambda_{ijk}$ couplings, which renders the $\lambda_{331}$ coupling to be zero.

It is useful to contrast the $LLE$ sector of RPV SUSY with the Zee model. In both setups, the relevant coupling is antisymmetric in the two $SU(2)_L$-doublet indices: $\lambda_{ijk}=-\lambda_{jik}$ in $\tfrac12\lambda_{ijk}\hat L_i\hat L_j\hat E_k^c$ and $f_{ij}=-f_{ji}$ in $f_{ij}L_i^T Ci\sigma_2L_jh^+$. However, the way this antisymmetry impacts $\varepsilon_{\tau\tau}$ differs once one attempts to mediate the NSI through an $SU(2)_L$-doublet scalar. In the Zee model, the coupling $f_{31}$ is allowed and directly connects the $\nu_\tau$ component of $L_3$ to the electron component of $L_1$, thereby generating an $e$--$\nu_\tau$ interaction and hence an $\varepsilon_{\tau\tau}$ contribution. In RPV SUSY, by contrast, the left-handed stau and the tau sneutrino belong to the same $SU(2)_L$-doublet superfield $\hat L_3$. A $\widetilde{\tau}_L$-mediated contribution to $\varepsilon_{\tau\tau}$ would require both $SU(2)_L$-doublet superfields in $\lambda_{ijk} \hat L_i\hat L_j\hat E_k^c$ to carry the third-generation label. The reason is that the relevant tree-level exchange involves two distinct $\hat L$ insertions: one must supply the scalar component $\widetilde{\tau}_L$ from $\hat L_3$, while the other must supply the fermionic component $\nu_\tau$ also from $\hat L_3$, forcing a coupling of the form $\lambda_{33k}$. This is precisely forbidden by $\lambda_{ijk}=-\lambda_{jik}$, implying $\lambda_{33k}=0$, and therefore the Zee-like $\widetilde{\tau}_L$ channel for $\varepsilon_{\tau\tau}$ is absent in the RPV3 realization.

In this sense, the crucial difference from the Zee model is not the presence or absence of antisymmetry itself, since both setups inherit an antisymmetric structure from the $SU(2)_L$ contraction. Rather, it is the way the mediator is embedded in the theory that matters: in RPV SUSY, $\widetilde{\tau}_L$ is part of an $SU(2)_L$-doublet superfield, so the antisymmetry forbids the would-be $\varepsilon_{\tau\tau}$ channel, whereas in the Zee model the charged scalar is an external singlet and no analogous obstruction arises.
\\
\\
{\bf (iv) Broader implications of UV-imposed correlations}

Taken together, the comparison above underlines a broader message: although RPV SUSY admits a rich phenomenology, its symmetry-driven microscopic structure typically correlates interactions and thereby brings along additional, often nontrivial, phenomenological constraints; the detailed patterns of viable low-energy effects are therefore controlled not only by the operator basis itself, but also by UV-imposed correlations among coefficients and the accompanying constraints. In the $\widetilde{\tau}_R$-mediated cases, the interactions responsible for the NSI of interest are typically accompanied by correlated (stringent) charged-lepton-flavor and universality constraints that can become unavoidable and strongly restrict the parameter space. In the $\widetilde{\tau}_L$-mediated cases, the supersymmetric embedding of the $SU(2)_L$ structure in the superpotential enforces additional relations among vertices; in particular, the antisymmetry $\lambda_{ijk}=-\lambda_{jik}$ can eliminate seemingly useful couplings and thereby remove Zee-like channels altogether. These two representative cases highlight how different UV completions that look similar at low energies can nonetheless lead to distinct patterns and constraints, offering potential handles to discriminate among underlying new physics scenarios.

\section{Conclusions} \label{sec:conclusions}

In this work, we examined neutrino nonstandard interactions in matter induced by the trilinear $LLE$ sector of $R$-parity-violating supersymmetry in a third-generation-dominated (RPV3-like) spectrum. We provided a systematic organization of the relevant coupling patterns, established their mapping onto the standard NSI parametrization, and quantified the ranges of NSI that remain viable under current experimental constraints.

Our phenomenological analysis combines complementary information from colliders, precision charged-lepton observables, and neutrino measurements. In particular, we recast existing LHC searches for third-generation sleptons for the spectra relevant to our setup, and we incorporate LEP, lepton-decay/universality, lepton-flavor-violating, and neutrino-scattering/oscillation (including global-fit) constraints, together with an illustration of the prospective reach of future long-baseline facilities. Within our working setup, percent-level NSI is possible in some channels. The largest magnitudes allowed by current constraints are
$\varepsilon_{\mu\mu}=-0.56\%$ for $\lambda_{123}\neq 0$,
$\varepsilon_{ee}=8\%$ for $\lambda_{131}\neq 0$,
$\varepsilon_{\tau\tau}=-0.25\%$ for $\lambda_{133}\neq 0$,
$\varepsilon_{\mu\mu}=3.8\%$ for $\lambda_{231}\neq 0$,
and $\varepsilon_{\mu\tau}=-0.29\%$ for $\lambda_{123},\lambda_{133}\neq 0$.

We also compared the resulting NSI patterns with those in the Zee model, emphasizing that the supersymmetric $LLE$ origin leads to characteristic correlations among vertices and constraints (and hence distinct NSI textures) even when the low-energy operator structures appear similar. Looking ahead, improved oscillation and scattering data, combined with direct information on third-generation sleptons, can provide a useful handle to discriminate among different microscopic origins of NSI.

Overall, our results furnish a concrete benchmark for $LLE$-induced NSI in an RPV3-like spectrum and delineate the maximal NSI sizes compatible with current bounds.

\acknowledgments
The author is grateful to Bhupal Dev and Amarjit Soni for valuable discussions and for collaboration at an early stage of this project. This work was partially supported by the U.S. Department of Energy under Grant No.~DE-SC0017987, and by the National Science Foundations of China under Grants No.~12275052 and No.~12147101.

\appendix
\section{Details of the LHC Slepton-Search Recast}
\label{sec:validation}

In this appendix, we summarize the procedure used for the LHC slepton-search recast discussed in Sec.~\ref{sec:LHC constraints}. We first check our implementation against the direct-slepton simplified model considered in Ref.~\cite{ATLAS:2019lff}. We then apply the same event selection to the RPV-$LLE$ scenarios studied in this work.

We use the ATLAS SR-SF-0J signal region of Ref.~\cite{ATLAS:2019lff}, which targets same-flavor opposite-sign dilepton events with zero non-$b$-tagged jets. In our implementation, we impose the corresponding same-flavor event selection, including two opposite-sign electrons or muons, the dilepton invariant-mass requirement $m_{\ell\ell}>121.2~\mathrm{GeV}$, the missing-transverse-energy requirement $E_T^{\rm miss}>110~\mathrm{GeV}$, and the inclusive $m_{T2}>100~\mathrm{GeV}$ selection. The signal events are generated with \texttt{MadGraph5\_aMC@NLO}~\cite{Alwall:2014hca}. The resulting visible signal yield is then compared with the published 95\% CL limit for this signal region.

As a check of the recast procedure, we consider the direct-slepton simplified-model topology in Fig.~7(c) of Ref.~\cite{ATLAS:2019lff},
\begin{align}
pp\to \widetilde{\ell}^{+}\widetilde{\ell}^{-}~,\quad \widetilde{\ell}^{\pm}\to \ell^{\pm}\widetilde{\chi}_1^0~,
\end{align}
with $\widetilde{\ell}=\widetilde e,\widetilde\mu$ and ${\rm BR}(\widetilde{\ell}\to \ell\widetilde{\chi}_1^0)=100\%$, since this benchmark is the closest ATLAS simplified-model topology to the stau-pair production and decay topologies relevant for our RPV-$LLE$ reinterpretation. To quantify the agreement, we compare the slepton mass reach obtained from our SR-SF-0J implementation with the ATLAS exclusion contour at several representative LSP masses, $m_{\widetilde{\chi}_1^0}=0,100,200,300~\mathrm{GeV}$, as summarized in Table~\ref{tab:LHC_validation}.

\begin{table}[htbp]
\centering
\begin{tabular}{c|c|c}
\hline
$m_{\widetilde{\chi}_1^0}$ [GeV] 
& ATLAS reach [GeV] 
& This recast [GeV] \\
\hline
0   & $\sim 700$ & 639 \\
100 & $\sim 700$ & 638 \\
200 & $\sim 690$ & 637 \\
300 & $\sim 690$ & 635 \\
\hline
\end{tabular}
\caption{Comparison between the ATLAS direct-slepton mass reach in the simplified-model topology of Fig.~7(c) of Ref.~\cite{ATLAS:2019lff} and our SR-SF-0J recast implementation. The quoted ATLAS values are read from the corresponding exclusion contour.}
\label{tab:LHC_validation}
\end{table}

The resulting mass reach differs from the ATLAS contour by about $9\%$, which represents a reasonable level of agreement for a recast based on the published SR-SF-0J information. The remaining difference can naturally arise from the NLO+NLL normalization of the slepton-pair production cross section and the full statistical combination of exclusive signal and control regions used in the ATLAS result~\cite{ATLAS:2019lff}. As a result, our SR-SF-0J recast gives a smaller effective signal normalization and hence a slightly weaker mass reach. The comparison also shows a mild dependence on the LSP mass in the range $m_{\widetilde{\chi}_1^0}=0$--$300~\mathrm{GeV}$: both the ATLAS contour and our recast give an almost constant slepton mass reach. This is because the mass splitting remains large in this region, so that the charged leptons and missing transverse momentum are sufficiently hard for the SR-SF-0J selection.

\begin{figure*}[htbp]
    \centering
    {\includegraphics[width=8cm]{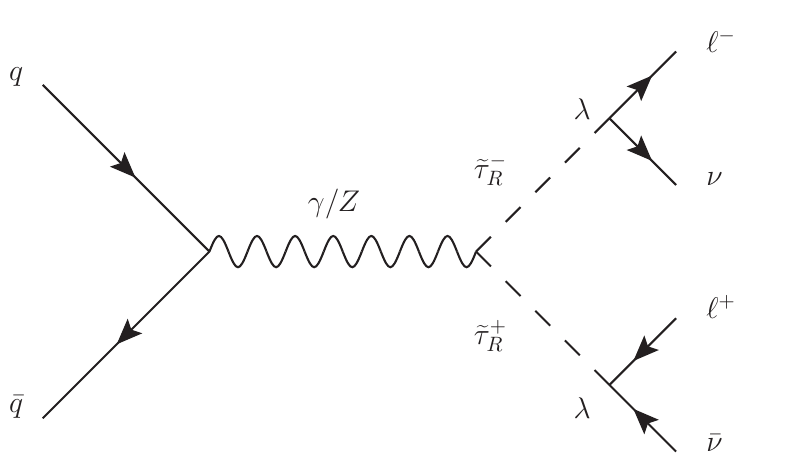}}
    {\includegraphics[width=8cm]{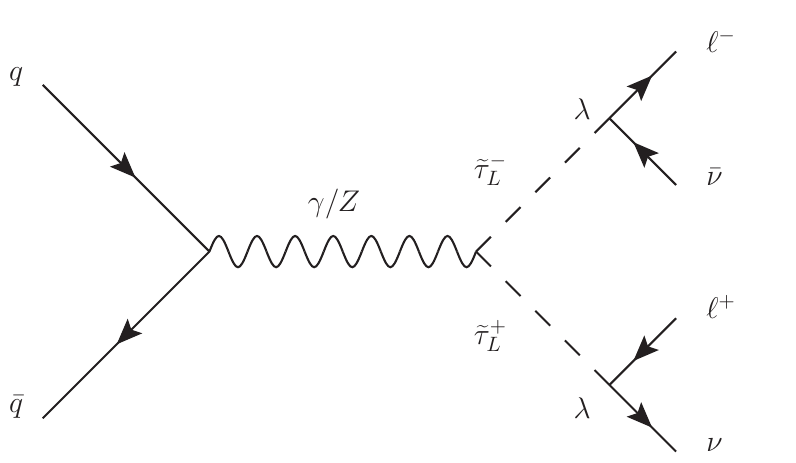}}
    \caption{Schematic $\widetilde{\tau}_R\widetilde{\tau}_R$ and $\widetilde{\tau}_L\widetilde{\tau}_L$ pair-production and decay topologies illustrating the channels entering the LHC recast. Here $\ell=e,\mu,\tau$, and the lepton and neutrino flavors depend on the relevant $LLE$ coupling assignment in each case. If $\ell=\tau$, subsequent leptonic $\tau$ decays can also contribute to the visible $ee+\slashed E_T$ or $\mu\mu+\slashed E_T$ final states entering the SR-SF-0J analysis~\cite{ATLAS:2019lff}.}
    \label{fig:LHC}
\end{figure*}

We then apply the same recast procedure to the RPV-$LLE$ scenarios considered in the main text. The corresponding schematic $\widetilde{\tau}_R\widetilde{\tau}_R$ and $\widetilde{\tau}_L\widetilde{\tau}_L$ pair-production topologies are illustrated in Fig.~\ref{fig:LHC}. Here $\ell$ and $\nu$ denote generic charged leptons and neutrinos, whose flavors depend on the relevant $LLE$ coupling assignment in each case. We work in the limit where the neutralino is decoupled from the signal topology, and generate the relevant $\widetilde{\tau}_R\widetilde{\tau}_R$ and/or $\widetilde{\tau}_L\widetilde{\tau}_L$ pair-production events for each coupling assignment. The staus subsequently decay through the corresponding $LLE$ interactions. Since the ATLAS SR-SF-0J region is defined in terms of same-flavor opposite-sign electrons or muons, we include all channels that populate the visible $ee+\slashed E_T$ or $\mu\mu+\slashed E_T$ final states after event selection. This includes channels where the stau decays directly produce electrons or muons, as well as channels in which one or both staus decay into $\tau$ leptons followed by leptonic $\tau$ decays.

For each RPV-$LLE$ case, we sum over the relevant $\widetilde{\tau}_R\widetilde{\tau}_R$ and/or $\widetilde{\tau}_L\widetilde{\tau}_L$ production and decay channels contributing to the SR-SF-0J topology. The resulting signal yield is compared with the observed 95\% CL limit in SR-SF-0J. This leads to the black-shaded exclusion regions in the $(m_{\widetilde{\tau}},\lambda)$ plane shown in Fig.~\ref{fig:Results}. The resulting lower bound is approximately $m_{\widetilde{\tau}}\gtrsim 350$--$550~\mathrm{GeV}$ at 95\% CL from the 13~TeV LHC dataset with an integrated luminosity of $139~\mathrm{fb}^{-1}$, with the value depending on the case under consideration.

\bibliographystyle{apsrev4-1}
\bibliography{references}

\end{document}